\documentclass[aps,prl,twocolumn,superscriptaddress,showpacs,floatfix,amsmath,amssymb]{revtex4-1}
\usepackage{graphicx,xcolor}
\usepackage{mathtools}
\usepackage{bm}

\begin{document}

\title{On the Nature of Localization in Ti doped Si}
 
\author{Yi Zhang}
\email{zhangyiphys@gmail.com}
\affiliation{Department of Physics \& Astronomy, Louisiana State University, Baton Rouge, Louisiana 70803, USA}
\affiliation{Center for Computation \& Technology, Louisiana State University, Baton Rouge, Louisiana 70803, USA}
\author{R. Nelson}
\affiliation{Institute of Inorganic Chemistry, RWTH Aachen University, Landoltweg 1, 52056 Aachen, Germany}
\author{K.-M.\ Tam}
\affiliation{Department of Physics \& Astronomy, Louisiana State University, Baton Rouge, Louisiana 70803, USA}
\affiliation{Center for Computation \& Technology, Louisiana State University, Baton Rouge, Louisiana 70803, USA}
\author{W.\ Ku}
\affiliation{Department of Physics and Astronomy, Shanghai Jiao Tong University, Shanghai 200240, China}
\author{U.\ Yu}
\affiliation{Department of Physics and Photon Science, GIST, Gwangju 61005, South Korea}
\author{N. S. Vidhyadhiraja} 
\affiliation{Theoretical Sciences Unit, 
Jawaharlal Nehru Centre for Advanced 
Scientific Research, Bangalore-560064, India}
\author{H. Terletska}
\affiliation{Department of Physics and Astronomy, Middle Tennessee State University, Murfreesboro, TN 37132, USA}
\author{J.\ Moreno}
\affiliation{Department of Physics \& Astronomy, Louisiana State University, Baton Rouge, Louisiana 70803, USA}
\affiliation{Center for Computation \& Technology, Louisiana State University, Baton Rouge, Louisiana 70803, USA}
\author{M.\ Jarrell}
\email{jarrellphysics@gmail.com}
\affiliation{Department of Physics \& Astronomy, Louisiana State University, Baton Rouge, Louisiana 70803, USA}
\affiliation{Center for Computation \& Technology, Louisiana State University, Baton Rouge, Louisiana 70803, USA}
\author{T.\ Berlijn}
\email{tberlijn@gmail.com}
\affiliation{Center for Nanophase Materials Sciences, Oak Ridge National Laboratory, Oak Ridge, TN 37831, USA}
\affiliation{Computer Science and Mathematics Division, Oak Ridge National Laboratory, Oak Ridge, Tennessee 37831, USA}

\begin{abstract}
Intermediate band semiconductors hold the promise to significantly improve the efficiency of solar cells, but only if the intermediate impurity band is metallic. We apply a recently developed first principles method to investigate the origin of electron localization in Ti doped Si, a promising candidate for intermediate band solar cells. Although Anderson localization is often overlooked in the context of intermediate band solar cells, our results show that in Ti doped Si it plays a more important role in the metal insulator transition than Mott localization. Implications for the theory of intermediate band solar cells are discussed.
\end{abstract}


\maketitle
 
\begin{figure}[h!]
 \includegraphics[trim = 0mm 0mm 0mm 0mm,width=0.7\columnwidth,clip=true]{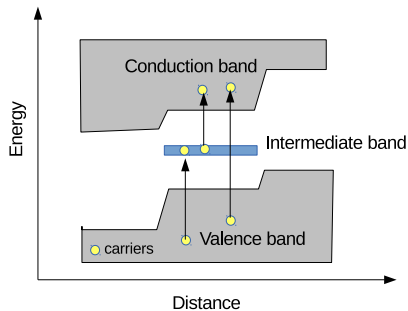}
 \caption{(color online) Schematic of an intermediate band solar cell (adapted from~\cite{A_Luque_2012}). Intermediate band states can dramatically improve the efficiency of solar cells by enabling two-photon processes which leads to the increase of photocurrent in a system.}
 \label{fig:IBSC}
\end{figure}

\textit{Introduction.}---Intermediate-band solar cells (IBSCs) have been proposed as a candidate for the third generation of photovoltaics~\cite{A_Luque_1997,A_Luque_2010,Y_Okada_2015}. Unlike  conventional solar cell materials, intermediate-band photovoltaics are doped with deep-level impurities that induce a partially filled intermediate band located between the valence and the conduction band as shown in Fig.~\ref{fig:IBSC}. This provides an extra channel for the promotion of an electron from the valence to the conduction band by absorbing two low energy photons instead of one photon with energy greater than the band gap. The extra two-photon channel leads to an increase of photocurrent without decreasing the photovoltage, which could greatly enhance the efficiency of solar cells~\cite{A_Luque_1997}. 

However, the deep-level impurity band also introduces electron-hole pair recombination centers, which normally lead to the increase of nonradiative Shockley-Reed-Hall (SRH) processes\cite{Shockley_1952,Hall_1952} that are detrimental to the efficiency of the solar cell.
When an electron or hole is captured by a deep-level impurity state the change in charge around the impurity causes local atomic displacements. According to the microscopic theory of Lang and Henry~\cite{lang_1975} these in turn strongly increase the capture cross-section of excited conduction electrons and valence holes into the intermediate band. Based on this theory, Luque \textit{et al.}~\cite{A_Luque_2006} argued that if the intermediate band becomes delocalized due to a large density of impurities, the charge of the trapped electron or hole will spread out. This in turn could suppress the atomic displacements and therefore the nonradiative recombinations. Consequently, a central question is how many impurities are needed to induce an insulator-metal transition in the intermediate band. From a general perspective this question is not only relevant for the efficiency of intermediate band solar cells, but is in fact a fundamental question in condensed matter physics. 

In 1977 Anderson and Mott shared one third each of the Nobel prize in physics in part for their study of the localization of electrons in semiconductors. Although they shared this Nobel prize they each had a distinct argument why the electrons become localized~\cite{p_anderson_1958,Mott_1968}. In Mott's model the localization of electrons, or rather the lack thereof, is controlled by the screening of the impurity potentials due to the long-range Coulomb interaction. When an impurity is isolated, it tightly traps the doped carriers. However, when the impurity concentration increases the electrons from one impurity screen the potential of a neighboring impurity thereby causing the electrons to be delocalized. We note here that Mott localization should not be confused with Mott-Hubbard localization~\cite{imada_1998}, in which intra-atomic Coulomb repulsion causes localization by opening a Mott gap. In Anderson's model the localization of electrons occurs purely due to the impurities being disordered. Most studies on IBSCs consider only Mott's criterion for localization~\cite{Marti_2008,A_Luque_2010,A_Luque_2012,J_Olea_2008,J_Olea_2009,J_Olea_2010,J_Olea_2012,K_Sanchez_2009,Gonzales_2009,Winkler_2011,zhou_2013_1,zhou_2013_2,Pastor_2013,Hu_2014,Dong_2015,Y_Okada_2015,Flores_2016,Liu_2016}. On the other hand Anderson localization in the context of IBSCs is examined less, either via approximate models~\cite{A_Luque_2006} or phenomenological fits~\cite{Winkler_2011} and rarely via first principles calculations~\cite{Carnio_2017}. Unbiased first principles calculations that take into account the material specifics can provide a unique perspective to investigate the relative importance of these two localization mechanisms in IBSCs. 

Among the intermediate band semiconductors, Si doped with elements such as Ti has the clear advantage that the host semiconductor is well studied.  Moreover, experimental indications for the promise of Ti doped Si are found in electrical resistivity and carrier lifetime measurements~\cite{J_Olea_2008,J_Olea_2010}. However, to reach an insulator-metal transition in the intermediate band, Ti concentrations beyond the solubility regime are required and non-equilibrium crystal growing techniques need to be applied, which are challenging~\cite{E_Antolin_2009}. Therefore independent first principles simulations including the effects of disorder will provide valuable guidance towards achieving high efficiency in Ti doped Si-based IBSCs. 

In this letter, we systematically study the metal-insulator transition in Ti doped Si as a function of Ti concentration, by combining two recently developed techniques, the Effective Disorder Hamiltonian Method (EDHM)~\cite{Berlijn_2011}  and the Typical Medium Dynamical Cluster Approximation (TMDCA)~\cite{c_ekuma_14}. We explore the mobility edge separating the delocalized and localized electron states in  the intermediate band, and find the critical impurity concentration of the localization transition. Moreover, by theoretically separating the effect of Mott and Anderson localization, we are able to compare these two mechanisms, and find that Anderson dominates over Mott localization in Ti doped Si.

\textit{Methods.}---First principles simulations take into account the multi-orbital nature of materials and the complex non-local structure of realistic impurity potentials. However, Anderson localization is usually not investigated from first-principles because localized states can be very large and typically need to be simulated with hundreds of thousands of lattice sites~\cite{vas_2008}. To overcome the computational expense we have recently developed a method that combines the EDHM and the TMDCA to study Anderson localization from first principles~\cite{y_zhang_15}. We have already applied this combined method to superconductors~\cite{y_zhang_15}, dilute magnetic semiconductors~\cite{y_zhang_17}, and here are applying it to the intermediate band semiconductor Ti doped Si. For another recent computational approach to study Anderson localization from first principles we refer to Ref.~\onlinecite{Carnio_2017}.    

The EDHM~\cite{Berlijn_2011} is a Wannier function~\cite{marzari97_prb56_12847,w_ku_02} based method which allows to derive low-energy tight-binding models of disordered materials from DFT calculations. Specifically models of both undoped Si and a supercell with a single Ti impurity are derived in the Wannier basis functions of Si-$s$, Si-$p$ and Ti-$d$, and the impurity potential is captured by the difference of these two models. Experimental measurements and theoretical calculations \cite{J_Olea_2008,K_Sanchez_2009} have shown that the Ti dopants are mostly interstitial impurities rather than (Si,Ti) substitutions and hence we focus here on Ti interstitials. To capture the experimental band-gap of Si we apply the LDA+U approximation, which we found to compare accurately with the modified Becke-Johnson potential~\cite{supp,tran_09}. In this study we used three different sizes of supercells: TiSi$_{8}$, TiSi$_{64}$ and TiSi$_{216}$ which lead to three different impurity potentials.  

Next we use the low-energy tight-binding model of pure Si and the Ti impurity potentials obtained from the EDHM as input for the TMDCA. The TMDCA is a cluster extension of the typical medium theory (TMT)~\cite{v_dobrosavljevic_03}, which in turn is a modification of the coherent potential approximation (CPA)~\cite{p_soven_67}, where a geometric average of the local density of states (DOS): $($DOS$_1\cdot$DOS$_2\cdot...\cdot$DOS$_N)^{1/N}$ is carried out in the impurity solver instead of the usual arithmetic average: $($DOS$_1+$DOS$_2+...+$DOS$_N)/N$. Here DOS$_i$ is the DOS at a particular site in a particular disorder configuration and $N$ is the total number of sites. The resulting geometrically averaged DOS or typical density of states (TDOS) captures the physics of localization~\cite{v_dobrosavljevic_03,kpm}. TDOS is finite in the delocalized phase and vanishes at the localized phase and so serves as an order parameter for the transition. Therefore, by comparing DOS and TDOS in the same plot, we are able to determine which states are localized and which are metallic.  TMDCA overcomes the restrictions of the TMT and accurately predicts the critical disorder strength of the single-band Anderson model with uniform disorder \cite{c_ekuma_14}.  In order to deal with more complicated realistic systems, the TMDCA is extended to systems with off-diagonal disorder \cite{h_ter_14} and to multi-band systems \cite{y_zhang_15}.

\begin{figure}[t!]
 \includegraphics[trim = 0mm 0mm 0mm 0mm,width=1\columnwidth,clip=true]{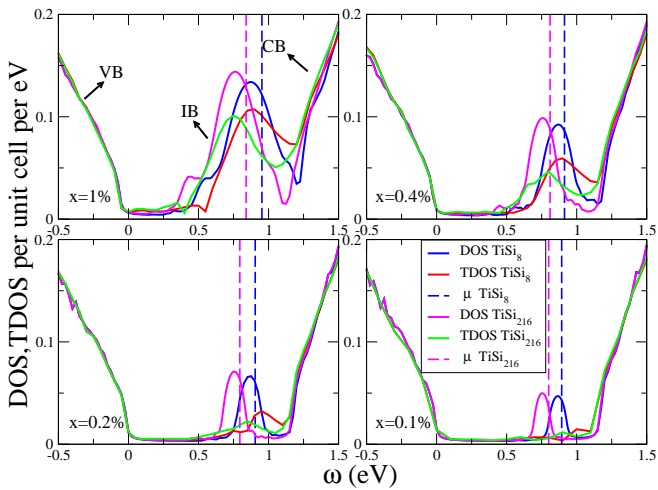}
 \caption{(color online) Density of States (DOS) and Typical Density of States (TDOS) of  Ti doped Si for various Ti concentrations: $x=$1\%, 0.4\%, 0.2\%, 0.1\%. Two sets of results are presented based on the impurity potentials from  supercell calculations with two difference sizes: TiSi$_8$ and TiSi$_{216}$. VB, CB, and IB correspond to the valence, conduction and intermediate band, respectively. The chemical potentials are indicated by dashed lines.}
 \label{fig:dos_tdos_mu_x_np}
\end{figure}

\textit{Results.}---First, we derive the critical concentration for the metal insulator transition in Ti doped Si by calculating DOS and TDOS for various Ti concentrations, $x$. We have checked convergence against various computational parameters~\cite{supp}. Fig.~\ref{fig:dos_tdos_mu_x_np} displays the concentration $x$ evolution of the DOS and TDOS. The band roughly above 1.25~eV corresponds to the conduction band and the one  below 0~eV is the valence band. The partially filled intermediate band appears between these two energies. Let us focus first on the results derived from the TiSi$_{216}$ supercell. For the relatively large Ti concentrations, $x$=1\%, the TDOS of the impurity band is finite indicating that its states are delocalized, i.e., metallic. As the Ti concentration $x$ decreases, the TDOS of the intermediate band gradually decreases and starts to vanish at concentrations between $x$=0.2\% and $x$=0.1\% signaling the localization transition. These values correspond to a critical Ti concentration between 1.0$\times$10$^{20}$ cm$^{-3}$ and 5.0$\times$10$^{19}$ cm$^{-3}$, which is consistent with the available experimental results~\cite{J_Olea_2012}. We have checked that neither lattice relaxation, nor spin-polarization effects change this conclusion significantly~\cite{supp}.  Both theoretical calculations and experiments have shown that such high concentrations of Ti in Si are thermodynamically unstable~\cite{E_Antolin_2009,K_Sanchez_2009}. Hence non-equilibrium growth techniques have been employed to increase doping~\cite{E_Antolin_2009}. Drawbacks of such preparation methods are  inhomogeneous distribution of dopant and damage to the crystal structure. Therefore our first principles derivation of the critical concentration is a valuable benchmark. However, the theoretical derivation of the critical concentration by itself does not answer the question what causes the metal-insulator transition in Ti doped Si: is it Mott localization or Anderson localization?

To investigate the relative importance of Mott's and Anderson's localization mechanisms we will now explore the effects of screening in our simulation. In Mott's original picture~\cite{Mott_1968}, the electronic impurity states are assumed to be localized, discrete, and bound to the impurity.  As the number of impurities increases, however, the binding potential of one impurity undergoes Thomas-Fermi screening by the long-range Coulomb potentials of the electrons on the surrounding impurities. The Mott transition from insulator to metal occurs when this screening reduces the strength of the impurity potential below a critical value, squeezing the impurity state into the continuum and forming a metal. Unlike the effects of Mott-Hubbard localization caused by intra-atomic Coulomb repulsion, Mott's model based on Thomas-Fermi screening can be captured accurately within DFT. In doped semiconductors Mott and Anderson localization are entangled~\cite{Rosenbaum_1983} and it is usually quite challenging to distinguish them. However, it turns out that within our EDHM+TMDCA method the separation of Mott's and Anderson's mechanisms is natural. 
 
In Mott's picture of localization, the states are pushed into the continuum due to the screening of the potential, while in Anderson localization, the states are localized due to disorder. Therefore, by tuning the strength of screening and disorder separately, we are able to distinguish the effect of Mott and Anderson localizations. In our method, the strength of disorder is tuned by the concentration of impurities in the TMDCA calculation, while the screening effect as captured by the EDHM is frozen in the impurity potential. By changing the size of the supercell used for the EDHM when deriving the impurity potential, we have a separate knob to tune the strength of the screening effect. Based on this, we derive the impurity potential from three different supercell sizes: TiSi$_8$, TiSi$_{64}$ and TiSi$_{216}$. Given that the Ti concentration in the TiSi$_8$ supercell is 27 times larger than in the TiSi$_{216}$ supercell one would expect based on Mott's mechanism a strong reduction of the impurity potential and therefore a decrease in the localization. However,  as shown in Fig.~\ref{fig:dos_tdos_mu_x_np} we distinguish no significant effect on the localization from the TMDCA based on these two impurity potentials. For each of the four disorder concentrations we see only minor changes in the DOS and TDOS for the TiSi$_8$ and TiSi$_{216}$ derived impurity potentials. The relative difference between the DOS and TDOS is much more sensitive to changes in disorder than to the changes in the screening. The DOS and TDOS vary even much less when the impurity potentials from TiSi$_{64}$ and TiSi$_{216}$ are compared~\cite{supp}. More importantly, the critical impurity concentration for all three investigated screening strengths lies between $x$=0.2\% and $x$=0.1\%. This indicates that the screening induced Mott localization plays a marginal role here compared to Anderson localization, despite the fact that most studies on IBSCs focus on Mott's criterion only. 

\begin{figure}[h!]
 \includegraphics[trim = 0mm 0mm 0mm 0mm,width=1\columnwidth,clip=true]{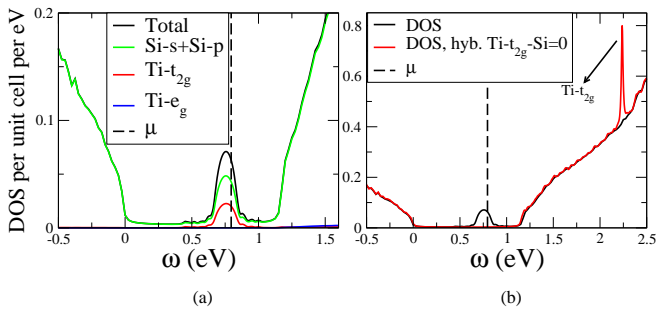}
 \caption{(color online) Density of States (DOS) of Ti doped Si for Ti concentration $x=$0.2\% based on the impurity potential derived from the TiSi$_{216}$ supercell.  (a) Orbital resolved contributions.  (b) DOS when hybridization between Ti-$t_{2g}$ and Si-$s$, Si-$p$ is removed.  The dashed line indicates the chemical potential. }
 \label{fig:dos_remove_d}
\end{figure}

To understand the weak effect of Mott's screening on Ti doped Si we take a closer look at the electronic structure of the impurity band complex. Fig.~\ref{fig:dos_remove_d}(a) shows the DOS of Si with 0.2\% of Ti impurities, now resolving the partial contributions from Ti-$t_{2g}$, Ti-$e_{g}$ and Si-$s$+Si-$p$. As we can see the intermediate band complex consists of a strong mixture of Ti-$t_{2g}$, Si-$s$ and Si-$p$. Clearly the hybridization of the Ti-$t_{2g}$ orbitals with Si-$s$/Si-$p$  plays an important role in the formation of the impurity band. To better illustrate this we plot in Fig. \ref{fig:dos_remove_d}(b) the total DOS for a calculation in which we switch off the hybridization of Ti-$t_{2g}$ with Si-$s$ and Si-$p$ in our effective tight-binding model. Fig. \ref{fig:dos_remove_d}(b) shows that in that case the impurity band vanishes from the gap and ends up about 1~eV above the bottom of the conduction band. In other words the hybridization of Ti-$t_{2g}$ with Si-$s$ and Si-$p$ is what creates the impurity band and this explains why the effects of screening are so weak in Ti doped Si. The main effect of the Ti impurity is coming from the overlaps of the Ti-$t_{2g}$ wave functions with those of the Si-$s$ and Si-$p$ wave functions and those are affected only weakly by screening at most. For example, the largest element in our first principles derived impurity potential is a hopping element between Ti-$t_{2g}$ and a nearest neighboring Si-$p$ orbital. Its value of 1.4 eV differs only by 1~meV when its extracted from the TiSi$_{8}$ supercell instead of the TiSi$_{216}$ supercell. Based on this microscopic insight we expect that our conclusion on the weakness of screening effects in Ti doped Si can be generalized to other IB semiconductors. In particular, in transition metal doped intermediate band semiconductors such as Co doped Si~\cite{zhou_2013_2}, V doped In$_2$S$_3$~\cite{Raquel_2008}, Ti doped GaAs~\cite{Brandt_1989} and Cr doped AlP~\cite{Olsson_2009}, we can expect a strong hopping disorder, given that the transition metal $d$ impurity orbitals are highly distinct from the $s$ and $p$ host orbitals. On the other hand, in S doped Si the impurity and host atoms are chemically close to each other because S and Si are in the same row and only two columns apart in the periodic table. Therefore the impurity band in this case is expected to be less controlled by hopping to impurity sites and hence more susceptible to screening effects, explaining why long range Coulomb effects in S doped Si may play a more important role~\cite{Winkler_2011}.

Our finding that in Ti doped Si Anderson localization dominates over Mott localization has important consequences for the theory of intermediate band solar cells in this system and others like it, given that the nature of these two localization mechanisms is fundamentally different. First of all, the Mott transition is believed to be first order~\cite{Mott_1968}, whereas the Anderson transition is a second order phase transition~\cite{abrahams_2014}. Therefore, one expects a less abrupt lifetime recovery as a function of Ti doping for Anderson localization than for the Mott's mechanism. Furthermore, a Mott localized state is trapped by a single impurity whereas the Anderson localized state is typically trapped by a cluster of impurities that has a large extent in space~\cite{Carnio_2017}. This means that the charge in an Anderson localized state will be more spread out and less likely to cause non-radiative recombinations than in a Mott localized state. Finally, the Anderson transition is a quantum phase transition only defined at zero temperature~\cite{abrahams_2014}, whereas the relevant temperature for IBSCs is room temperature. However, it has been shown that effects of the Anderson localization, such as  variable range hopping, extend to room temperature and beyond~\cite{chopra_1970,Clayton_2013,hapert_1973}. Moreover, even if an electron hops between Anderson localized states via  interaction with phonons~\cite{Ambe_1971}, an important question is how fast it will do so. If the time scale is larger or comparable to the carrier lifetime then the Anderson localization should still strongly affect the non-radiative recombination rate. Given that both variable range hopping and non-radiative recombinations are controlled by phonons, it is conceivable that their time scales be comparable. The above implications for the theory of IBSCs highlight the richness of the physics of Anderson localization and that of disordered materials in general.

In summary, by combining two recently developed theoretical techniques, the EDHM and the TMDCA, we investigate from first principles the metal-insulator transition in the promising intermediate-band photovoltaic material Ti doped Si. We systematically study the localization in the impurity band and find  that the impurity band electrons delocalize for a Ti concentration between $x$=0.1\% and $x$=0.2\%. Our calculation can be applied to other systems with intermediate bands providing guidance to make highly efficient IB solar cells. Moreover, our approach provides a systematic way to study the nature of the localization transition by separating the effects of Mott and Anderson localization. Our results show that in Ti doped Si, Anderson localization dominates over Mott localization, despite that most studies on intermediate band solar cells consider Mott's criterion for localization only. The reason for the weakness of Mott localization here is that the impurity band is induced by the hopping between Ti-$t_{2g}$ and Si-$s$/Si-$p$ orbitals, an effect that can not be diminished by screening. Given the fundamental differences between Mott and Anderson localization our finding has important implications for the theory of intermediate band solar cells. 

This letter is based upon work supported by the U.S. Department of Energy, Office of Science, Office of Basic Energy Sciences under Award Number DE-SC0017861. Work by TB was performed at the Center for Nanophase Materials Sciences, a DOE Office of Science user facility. This manuscript has been authored by UT-Battelle, LLC under Contract No. DE-AC05-00OR22725 with the U.S. Department of Energy. This work used the high performance computational resources provided by the Louisiana Optical Network Initiative (http://www.loni.org), and HPC@LSU computing. Furthermore, this research used resources of the National Energy Research Scientific Computing Center, a DOE Office of Science User Facility supported by the Office of Science of the U.S. DOE under Contract No. DE-AC02-05CH11231.
This work also used resources of the Oak Ridge Leadership Computing Facility at the Oak Ridge National Laboratory, which is supported by the Office of Science of the U.S. Department of Energy under Contract No. DE-AC05-00OR22725.


\begin{thebibliography}{50}%
\makeatletter
\providecommand \@ifxundefined [1]{%
 \@ifx{#1\undefined}
}%
\providecommand \@ifnum [1]{%
 \ifnum #1\expandafter \@firstoftwo
 \else \expandafter \@secondoftwo
 \fi
}%
\providecommand \@ifx [1]{%
 \ifx #1\expandafter \@firstoftwo
 \else \expandafter \@secondoftwo
 \fi
}%
\providecommand \natexlab [1]{#1}%
\providecommand \enquote  [1]{``#1''}%
\providecommand \bibnamefont  [1]{#1}%
\providecommand \bibfnamefont [1]{#1}%
\providecommand \citenamefont [1]{#1}%
\providecommand \href@noop [0]{\@secondoftwo}%
\providecommand \href [0]{\begingroup \@sanitize@url \@href}%
\providecommand \@href[1]{\@@startlink{#1}\@@href}%
\providecommand \@@href[1]{\endgroup#1\@@endlink}%
\providecommand \@sanitize@url [0]{\catcode `\\12\catcode `\$12\catcode
  `\&12\catcode `\#12\catcode `\^12\catcode `\_12\catcode `\%12\relax}%
\providecommand \@@startlink[1]{}%
\providecommand \@@endlink[0]{}%
\providecommand \url  [0]{\begingroup\@sanitize@url \@url }%
\providecommand \@url [1]{\endgroup\@href {#1}{\urlprefix }}%
\providecommand \urlprefix  [0]{URL }%
\providecommand \Eprint [0]{\href }%
\providecommand \doibase [0]{http://dx.doi.org/}%
\providecommand \selectlanguage [0]{\@gobble}%
\providecommand \bibinfo  [0]{\@secondoftwo}%
\providecommand \bibfield  [0]{\@secondoftwo}%
\providecommand \translation [1]{[#1]}%
\providecommand \BibitemOpen [0]{}%
\providecommand \bibitemStop [0]{}%
\providecommand \bibitemNoStop [0]{.\EOS\space}%
\providecommand \EOS [0]{\spacefactor3000\relax}%
\providecommand \BibitemShut  [1]{\csname bibitem#1\endcsname}%
\let\auto@bib@innerbib\@empty
\bibitem [{\citenamefont {Luque}\ and\ \citenamefont
  {Mart\'{\i}}(1997)}]{A_Luque_1997}%
  \BibitemOpen
  \bibfield  {author} {\bibinfo {author} {\bibfnamefont {A.}~\bibnamefont
  {Luque}}\ and\ \bibinfo {author} {\bibfnamefont {A.}~\bibnamefont
  {Mart\'{\i}}},\ }\href {\doibase 10.1103/PhysRevLett.78.5014} {\bibfield
  {journal} {\bibinfo  {journal} {Phys. Rev. Lett.}\ }\textbf {\bibinfo
  {volume} {78}},\ \bibinfo {pages} {5014} (\bibinfo {year}
  {1997})}\BibitemShut {NoStop}%
\bibitem [{\citenamefont {Luque}\ and\ \citenamefont
  {Mart\'{\i}}(2010)}]{A_Luque_2010}%
  \BibitemOpen
  \bibfield  {author} {\bibinfo {author} {\bibfnamefont {A.}~\bibnamefont
  {Luque}}\ and\ \bibinfo {author} {\bibfnamefont {A.}~\bibnamefont
  {Mart\'{\i}}},\ }\href {\doibase 10.1002/adma.200902388} {\bibfield
  {journal} {\bibinfo  {journal} {Adv. Mater.}\ }\textbf {\bibinfo {volume}
  {22}},\ \bibinfo {pages} {160} (\bibinfo {year} {2010})}\BibitemShut
  {NoStop}%
\bibitem [{\citenamefont {Okada}\ \emph {et~al.}(2015)\citenamefont {Okada},
  \citenamefont {Ekins-Daukes}, \citenamefont {Kita}, \citenamefont {Tamaki},
  \citenamefont {Yoshida}, \citenamefont {Pusch}, \citenamefont {Hess},
  \citenamefont {Phillips}, \citenamefont {Farrell}, \citenamefont {Yoshida},
  \citenamefont {Ahsan}, \citenamefont {Shoji}, \citenamefont {Sogabe},\ and\
  \citenamefont {Guillemoles}}]{Y_Okada_2015}%
  \BibitemOpen
  \bibfield  {author} {\bibinfo {author} {\bibfnamefont {Y.}~\bibnamefont
  {Okada}}, \bibinfo {author} {\bibfnamefont {N.~J.}\ \bibnamefont
  {Ekins-Daukes}}, \bibinfo {author} {\bibfnamefont {T.}~\bibnamefont {Kita}},
  \bibinfo {author} {\bibfnamefont {R.}~\bibnamefont {Tamaki}}, \bibinfo
  {author} {\bibfnamefont {M.}~\bibnamefont {Yoshida}}, \bibinfo {author}
  {\bibfnamefont {A.}~\bibnamefont {Pusch}}, \bibinfo {author} {\bibfnamefont
  {O.}~\bibnamefont {Hess}}, \bibinfo {author} {\bibfnamefont {C.~C.}\
  \bibnamefont {Phillips}}, \bibinfo {author} {\bibfnamefont {D.~J.}\
  \bibnamefont {Farrell}}, \bibinfo {author} {\bibfnamefont {K.}~\bibnamefont
  {Yoshida}}, \bibinfo {author} {\bibfnamefont {N.}~\bibnamefont {Ahsan}},
  \bibinfo {author} {\bibfnamefont {Y.}~\bibnamefont {Shoji}}, \bibinfo
  {author} {\bibfnamefont {T.}~\bibnamefont {Sogabe}}, \ and\ \bibinfo {author}
  {\bibfnamefont {J.-F.}\ \bibnamefont {Guillemoles}},\ }\href {\doibase
  10.1063/1.4916561} {\bibfield  {journal} {\bibinfo  {journal} {Appl. Phys.
  Rev.}\ }\textbf {\bibinfo {volume} {2}},\ \bibinfo {pages} {021302} (\bibinfo
  {year} {2015})}\BibitemShut {NoStop}%
\bibitem [{\citenamefont {Luque}\ \emph {et~al.}(2012)\citenamefont {Luque},
  \citenamefont {Mart\'{\i}},\ and\ \citenamefont {Stanley}}]{A_Luque_2012}%
  \BibitemOpen
  \bibfield  {author} {\bibinfo {author} {\bibfnamefont {A.}~\bibnamefont
  {Luque}}, \bibinfo {author} {\bibfnamefont {A.}~\bibnamefont {Mart\'{\i}}}, \
  and\ \bibinfo {author} {\bibfnamefont {C.}~\bibnamefont {Stanley}},\ }\href
  {\doibase doi:10.1038/nphoton.2012.1} {\bibfield  {journal} {\bibinfo
  {journal} {Nat. Photonics}\ }\textbf {\bibinfo {volume} {6}},\ \bibinfo
  {pages} {146} (\bibinfo {year} {2012})}\BibitemShut {NoStop}%
\bibitem [{\citenamefont {Shockley}\ and\ \citenamefont
  {Read}(1952)}]{Shockley_1952}%
  \BibitemOpen
  \bibfield  {author} {\bibinfo {author} {\bibfnamefont {W.}~\bibnamefont
  {Shockley}}\ and\ \bibinfo {author} {\bibfnamefont {W.~T.}\ \bibnamefont
  {Read}},\ }\href {\doibase 10.1103/PhysRev.87.835} {\bibfield  {journal}
  {\bibinfo  {journal} {Phys. Rev.}\ }\textbf {\bibinfo {volume} {87}},\
  \bibinfo {pages} {835} (\bibinfo {year} {1952})}\BibitemShut {NoStop}%
\bibitem [{\citenamefont {Hall}(1952)}]{Hall_1952}%
  \BibitemOpen
  \bibfield  {author} {\bibinfo {author} {\bibfnamefont {R.~N.}\ \bibnamefont
  {Hall}},\ }\href {\doibase 10.1103/PhysRev.87.387} {\bibfield  {journal}
  {\bibinfo  {journal} {Phys. Rev.}\ }\textbf {\bibinfo {volume} {87}},\
  \bibinfo {pages} {387} (\bibinfo {year} {1952})}\BibitemShut {NoStop}%
\bibitem [{\citenamefont {Lang}\ and\ \citenamefont {Henry}(1975)}]{lang_1975}%
  \BibitemOpen
  \bibfield  {author} {\bibinfo {author} {\bibfnamefont {D.~V.}\ \bibnamefont
  {Lang}}\ and\ \bibinfo {author} {\bibfnamefont {C.~H.}\ \bibnamefont
  {Henry}},\ }\href {\doibase 10.1103/PhysRevLett.35.1525} {\bibfield
  {journal} {\bibinfo  {journal} {Phys. Rev. Lett.}\ }\textbf {\bibinfo
  {volume} {35}},\ \bibinfo {pages} {1525} (\bibinfo {year}
  {1975})}\BibitemShut {NoStop}%
\bibitem [{\citenamefont {Luque}\ \emph {et~al.}(2006)\citenamefont {Luque},
  \citenamefont {Mart\'{\i}}, \citenamefont {Antol\'{\i}n},\ and\ \citenamefont
  {Tablero}}]{A_Luque_2006}%
  \BibitemOpen
  \bibfield  {author} {\bibinfo {author} {\bibfnamefont {A.}~\bibnamefont
  {Luque}}, \bibinfo {author} {\bibfnamefont {A.}~\bibnamefont {Mart\'{\i}}},
  \bibinfo {author} {\bibfnamefont {E.}~\bibnamefont {Antol\'{\i}n}}, \ and\
  \bibinfo {author} {\bibfnamefont {C.}~\bibnamefont {Tablero}},\ }\href
  {\doibase https://doi.org/10.1016/j.physb.2006.03.006} {\bibfield  {journal}
  {\bibinfo  {journal} {Physica B}\ }\textbf {\bibinfo {volume} {382}},\
  \bibinfo {pages} {320 } (\bibinfo {year} {2006})}\BibitemShut {NoStop}%
\bibitem [{\citenamefont {Anderson}(1958)}]{p_anderson_1958}%
  \BibitemOpen
  \bibfield  {author} {\bibinfo {author} {\bibfnamefont {P.~W.}\ \bibnamefont
  {Anderson}},\ }\href {\doibase 10.1103/PhysRev.109.1492} {\bibfield
  {journal} {\bibinfo  {journal} {Phys. Rev.}\ }\textbf {\bibinfo {volume}
  {109}},\ \bibinfo {pages} {1492} (\bibinfo {year} {1958})}\BibitemShut
  {NoStop}%
\bibitem [{\citenamefont {Mott}(1968)}]{Mott_1968}%
  \BibitemOpen
  \bibfield  {author} {\bibinfo {author} {\bibfnamefont {N.~F.}\ \bibnamefont
  {Mott}},\ }\href {\doibase 10.1103/RevModPhys.40.677} {\bibfield  {journal}
  {\bibinfo  {journal} {Rev. Mod. Phys.}\ }\textbf {\bibinfo {volume} {40}},\
  \bibinfo {pages} {677} (\bibinfo {year} {1968})}\BibitemShut {NoStop}%
\bibitem [{\citenamefont {Imada}\ \emph {et~al.}(1998)\citenamefont {Imada},
  \citenamefont {Fujimori},\ and\ \citenamefont {Tokura}}]{imada_1998}%
  \BibitemOpen
  \bibfield  {author} {\bibinfo {author} {\bibfnamefont {M.}~\bibnamefont
  {Imada}}, \bibinfo {author} {\bibfnamefont {A.}~\bibnamefont {Fujimori}}, \
  and\ \bibinfo {author} {\bibfnamefont {Y.}~\bibnamefont {Tokura}},\ }\href
  {\doibase 10.1103/RevModPhys.70.1039} {\bibfield  {journal} {\bibinfo
  {journal} {Rev. Mod. Phys.}\ }\textbf {\bibinfo {volume} {70}},\ \bibinfo
  {pages} {1039} (\bibinfo {year} {1998})}\BibitemShut {NoStop}%
\bibitem [{\citenamefont {Mart{\'i}}\ \emph {et~al.}(2008)\citenamefont
  {Mart{\'i}}, \citenamefont {Marr{\'o}n},\ and\ \citenamefont
  {Luque}}]{Marti_2008}%
  \BibitemOpen
  \bibfield  {author} {\bibinfo {author} {\bibfnamefont {A.}~\bibnamefont
  {Mart{\'i}}}, \bibinfo {author} {\bibfnamefont {D.~F.}\ \bibnamefont
  {Marr{\'o}n}}, \ and\ \bibinfo {author} {\bibfnamefont {A.}~\bibnamefont
  {Luque}},\ }\href@noop {} {\bibfield  {journal} {\bibinfo  {journal} {Journal
  of Applied Physics}\ }\textbf {\bibinfo {volume} {103}},\ \bibinfo {pages}
  {073706} (\bibinfo {year} {2008})}\BibitemShut {NoStop}%
\bibitem [{\citenamefont {Olea}\ \emph {et~al.}(2008)\citenamefont {Olea},
  \citenamefont {Toledano-Luque}, \citenamefont {Pastor}, \citenamefont
  {Gonz{\'a}lez-D{\'i}az},\ and\ \citenamefont {M{\'a}rtil}}]{J_Olea_2008}%
  \BibitemOpen
  \bibfield  {author} {\bibinfo {author} {\bibfnamefont {J.}~\bibnamefont
  {Olea}}, \bibinfo {author} {\bibfnamefont {M.}~\bibnamefont
  {Toledano-Luque}}, \bibinfo {author} {\bibfnamefont {D.}~\bibnamefont
  {Pastor}}, \bibinfo {author} {\bibfnamefont {G.}~\bibnamefont
  {Gonz{\'a}lez-D{\'i}az}}, \ and\ \bibinfo {author} {\bibfnamefont
  {I.}~\bibnamefont {M{\'a}rtil}},\ }\href {\doibase 10.1063/1.2949258}
  {\bibfield  {journal} {\bibinfo  {journal} {J. Appl. Phys.}\ }\textbf
  {\bibinfo {volume} {104}},\ \bibinfo {pages} {016105} (\bibinfo {year}
  {2008})}\BibitemShut {NoStop}%
\bibitem [{\citenamefont {Olea}\ \emph {et~al.}(2009)\citenamefont {Olea},
  \citenamefont {Gonz{\'a}lez-D{\'i}az}, \citenamefont {Pastor},\ and\
  \citenamefont {M{\'a}rtil}}]{J_Olea_2009}%
  \BibitemOpen
  \bibfield  {author} {\bibinfo {author} {\bibfnamefont {J.}~\bibnamefont
  {Olea}}, \bibinfo {author} {\bibfnamefont {G.}~\bibnamefont
  {Gonz{\'a}lez-D{\'i}az}}, \bibinfo {author} {\bibfnamefont {D.}~\bibnamefont
  {Pastor}}, \ and\ \bibinfo {author} {\bibfnamefont {I.}~\bibnamefont
  {M{\'a}rtil}},\ }\href@noop {} {\bibfield  {journal} {\bibinfo  {journal}
  {Journal of Physics D: Applied Physics}\ }\textbf {\bibinfo {volume} {42}},\
  \bibinfo {pages} {085110} (\bibinfo {year} {2009})}\BibitemShut {NoStop}%
\bibitem [{\citenamefont {Olea}\ \emph {et~al.}(2010)\citenamefont {Olea},
  \citenamefont {Toledano-Luque}, \citenamefont {Pastor}, \citenamefont
  {San-Andr{\'e}s}, \citenamefont {M{\'a}rtil},\ and\ \citenamefont
  {Gonz{\'a}lez-D{\'i}az}}]{J_Olea_2010}%
  \BibitemOpen
  \bibfield  {author} {\bibinfo {author} {\bibfnamefont {J.}~\bibnamefont
  {Olea}}, \bibinfo {author} {\bibfnamefont {M.}~\bibnamefont
  {Toledano-Luque}}, \bibinfo {author} {\bibfnamefont {D.}~\bibnamefont
  {Pastor}}, \bibinfo {author} {\bibfnamefont {E.}~\bibnamefont
  {San-Andr{\'e}s}}, \bibinfo {author} {\bibfnamefont {I.}~\bibnamefont
  {M{\'a}rtil}}, \ and\ \bibinfo {author} {\bibfnamefont {G.}~\bibnamefont
  {Gonz{\'a}lez-D{\'i}az}},\ }\href {\doibase 10.1063/1.3391274} {\bibfield
  {journal} {\bibinfo  {journal} {J. Appl. Phys.}\ }\textbf {\bibinfo {volume}
  {107}},\ \bibinfo {pages} {103524} (\bibinfo {year} {2010})}\BibitemShut
  {NoStop}%
\bibitem [{\citenamefont {Olea}\ \emph {et~al.}(2012)\citenamefont {Olea},
  \citenamefont {Pastor}, \citenamefont {Garc{\'i}a-Hemme}, \citenamefont
  {Garc{\'i}a-Hernansanz}, \citenamefont {del Prado}, \citenamefont
  {M{\'a}rtil},\ and\ \citenamefont {Gonz{\'a}lez-D{\'i}az}}]{J_Olea_2012}%
  \BibitemOpen
  \bibfield  {author} {\bibinfo {author} {\bibfnamefont {J.}~\bibnamefont
  {Olea}}, \bibinfo {author} {\bibfnamefont {D.}~\bibnamefont {Pastor}},
  \bibinfo {author} {\bibfnamefont {E.}~\bibnamefont {Garc{\'i}a-Hemme}},
  \bibinfo {author} {\bibfnamefont {R.}~\bibnamefont {Garc{\'i}a-Hernansanz}},
  \bibinfo {author} {\bibfnamefont {{\'A}.}~\bibnamefont {del Prado}}, \bibinfo
  {author} {\bibfnamefont {I.}~\bibnamefont {M{\'a}rtil}}, \ and\ \bibinfo
  {author} {\bibfnamefont {G.}~\bibnamefont {Gonz{\'a}lez-D{\'i}az}},\ }\href
  {\doibase https://doi.org/10.1016/j.tsf.2012.07.014} {\bibfield  {journal}
  {\bibinfo  {journal} {Thin Solid Films}\ }\textbf {\bibinfo {volume} {520}},\
  \bibinfo {pages} {6614 } (\bibinfo {year} {2012})}\BibitemShut {NoStop}%
\bibitem [{\citenamefont {S\'anchez}\ \emph {et~al.}(2009)\citenamefont
  {S\'anchez}, \citenamefont {Aguilera}, \citenamefont {Palacios},\ and\
  \citenamefont {Wahn\'on}}]{K_Sanchez_2009}%
  \BibitemOpen
  \bibfield  {author} {\bibinfo {author} {\bibfnamefont {K.}~\bibnamefont
  {S\'anchez}}, \bibinfo {author} {\bibfnamefont {I.}~\bibnamefont {Aguilera}},
  \bibinfo {author} {\bibfnamefont {P.}~\bibnamefont {Palacios}}, \ and\
  \bibinfo {author} {\bibfnamefont {P.}~\bibnamefont {Wahn\'on}},\ }\href
  {\doibase 10.1103/PhysRevB.79.165203} {\bibfield  {journal} {\bibinfo
  {journal} {Phys. Rev. B}\ }\textbf {\bibinfo {volume} {79}},\ \bibinfo
  {pages} {165203} (\bibinfo {year} {2009})}\BibitemShut {NoStop}%
\bibitem [{\citenamefont {Gonzalez-D{\'i}az}\ \emph {et~al.}(2009)\citenamefont
  {Gonzalez-D{\'i}az}, \citenamefont {Olea}, \citenamefont {M{\'a}rtil},
  \citenamefont {Pastor}, \citenamefont {Mart{\'i}}, \citenamefont
  {Antol{\'i}n},\ and\ \citenamefont {Luque}}]{Gonzales_2009}%
  \BibitemOpen
  \bibfield  {author} {\bibinfo {author} {\bibfnamefont {G.}~\bibnamefont
  {Gonzalez-D{\'i}az}}, \bibinfo {author} {\bibfnamefont {J.}~\bibnamefont
  {Olea}}, \bibinfo {author} {\bibfnamefont {I.}~\bibnamefont {M{\'a}rtil}},
  \bibinfo {author} {\bibfnamefont {D.}~\bibnamefont {Pastor}}, \bibinfo
  {author} {\bibfnamefont {A.}~\bibnamefont {Mart{\'i}}}, \bibinfo {author}
  {\bibfnamefont {E.}~\bibnamefont {Antol{\'i}n}}, \ and\ \bibinfo {author}
  {\bibfnamefont {A.}~\bibnamefont {Luque}},\ }\href {\doibase
  https://doi.org/10.1016/j.solmat.2009.05.014} {\bibfield  {journal} {\bibinfo
   {journal} {Solar Energy Materials and Solar Cells}\ }\textbf {\bibinfo
  {volume} {93}},\ \bibinfo {pages} {1668 } (\bibinfo {year}
  {2009})}\BibitemShut {NoStop}%
\bibitem [{\citenamefont {Winkler}\ \emph {et~al.}(2011)\citenamefont
  {Winkler}, \citenamefont {Recht}, \citenamefont {Sher}, \citenamefont {Said},
  \citenamefont {Mazur},\ and\ \citenamefont {Aziz}}]{Winkler_2011}%
  \BibitemOpen
  \bibfield  {author} {\bibinfo {author} {\bibfnamefont {M.~T.}\ \bibnamefont
  {Winkler}}, \bibinfo {author} {\bibfnamefont {D.}~\bibnamefont {Recht}},
  \bibinfo {author} {\bibfnamefont {M.-J.}\ \bibnamefont {Sher}}, \bibinfo
  {author} {\bibfnamefont {A.~J.}\ \bibnamefont {Said}}, \bibinfo {author}
  {\bibfnamefont {E.}~\bibnamefont {Mazur}}, \ and\ \bibinfo {author}
  {\bibfnamefont {M.~J.}\ \bibnamefont {Aziz}},\ }\href {\doibase
  10.1103/PhysRevLett.106.178701} {\bibfield  {journal} {\bibinfo  {journal}
  {Phys. Rev. Lett.}\ }\textbf {\bibinfo {volume} {106}},\ \bibinfo {pages}
  {178701} (\bibinfo {year} {2011})}\BibitemShut {NoStop}%
\bibitem [{\citenamefont {Zhou}\ \emph
  {et~al.}(2013{\natexlab{a}})\citenamefont {Zhou}, \citenamefont {Liu},
  \citenamefont {Zhu}, \citenamefont {Song},\ and\ \citenamefont
  {Zhang}}]{zhou_2013_1}%
  \BibitemOpen
  \bibfield  {author} {\bibinfo {author} {\bibfnamefont {Y.}~\bibnamefont
  {Zhou}}, \bibinfo {author} {\bibfnamefont {F.}~\bibnamefont {Liu}}, \bibinfo
  {author} {\bibfnamefont {M.}~\bibnamefont {Zhu}}, \bibinfo {author}
  {\bibfnamefont {X.}~\bibnamefont {Song}}, \ and\ \bibinfo {author}
  {\bibfnamefont {P.}~\bibnamefont {Zhang}},\ }\href@noop {} {\bibfield
  {journal} {\bibinfo  {journal} {Applied Physics Letters}\ }\textbf {\bibinfo
  {volume} {102}},\ \bibinfo {pages} {222106} (\bibinfo {year}
  {2013}{\natexlab{a}})}\BibitemShut {NoStop}%
\bibitem [{\citenamefont {Zhou}\ \emph
  {et~al.}(2013{\natexlab{b}})\citenamefont {Zhou}, \citenamefont {Liu},\ and\
  \citenamefont {Song}}]{zhou_2013_2}%
  \BibitemOpen
  \bibfield  {author} {\bibinfo {author} {\bibfnamefont {Y.}~\bibnamefont
  {Zhou}}, \bibinfo {author} {\bibfnamefont {F.}~\bibnamefont {Liu}}, \ and\
  \bibinfo {author} {\bibfnamefont {X.}~\bibnamefont {Song}},\ }\href {\doibase
  10.1063/1.4794818} {\bibfield  {journal} {\bibinfo  {journal} {Journal of
  Applied Physics}\ }\textbf {\bibinfo {volume} {113}},\ \bibinfo {pages}
  {103702} (\bibinfo {year} {2013}{\natexlab{b}})},\ \Eprint
  {http://arxiv.org/abs/https://doi.org/10.1063/1.4794818}
  {https://doi.org/10.1063/1.4794818} \BibitemShut {NoStop}%
\bibitem [{\citenamefont {Pastor}\ \emph {et~al.}(2013)\citenamefont {Pastor},
  \citenamefont {Olea}, \citenamefont {del Prado}, \citenamefont
  {Garc{\'i}a-Hemme}, \citenamefont {Garc{\'i}a-Hernansanz}, \citenamefont
  {M{\'a}rtil},\ and\ \citenamefont {Gonz{\'a}lez-D{\'i}az}}]{Pastor_2013}%
  \BibitemOpen
  \bibfield  {author} {\bibinfo {author} {\bibfnamefont {D.}~\bibnamefont
  {Pastor}}, \bibinfo {author} {\bibfnamefont {J.}~\bibnamefont {Olea}},
  \bibinfo {author} {\bibfnamefont {A.}~\bibnamefont {del Prado}}, \bibinfo
  {author} {\bibfnamefont {E.}~\bibnamefont {Garc{\'i}a-Hemme}}, \bibinfo
  {author} {\bibfnamefont {R.}~\bibnamefont {Garc{\'i}a-Hernansanz}}, \bibinfo
  {author} {\bibfnamefont {I.}~\bibnamefont {M{\'a}rtil}}, \ and\ \bibinfo
  {author} {\bibfnamefont {G.}~\bibnamefont {Gonz{\'a}lez-D{\'i}az}},\ }\href
  {http://stacks.iop.org/0022-3727/46/i=13/a=135108} {\bibfield  {journal}
  {\bibinfo  {journal} {Journal of Physics D: Applied Physics}\ }\textbf
  {\bibinfo {volume} {46}},\ \bibinfo {pages} {135108} (\bibinfo {year}
  {2013})}\BibitemShut {NoStop}%
\bibitem [{\citenamefont {Hu}\ \emph {et~al.}(2014)\citenamefont {Hu},
  \citenamefont {Han}, \citenamefont {Liang}, \citenamefont {Xing},\ and\
  \citenamefont {Lou}}]{Hu_2014}%
  \BibitemOpen
  \bibfield  {author} {\bibinfo {author} {\bibfnamefont {S.}~\bibnamefont
  {Hu}}, \bibinfo {author} {\bibfnamefont {P.}~\bibnamefont {Han}}, \bibinfo
  {author} {\bibfnamefont {P.}~\bibnamefont {Liang}}, \bibinfo {author}
  {\bibfnamefont {Y.}~\bibnamefont {Xing}}, \ and\ \bibinfo {author}
  {\bibfnamefont {S.}~\bibnamefont {Lou}},\ }\href {\doibase
  https://doi.org/10.1016/j.mssp.2013.09.001} {\bibfield  {journal} {\bibinfo
  {journal} {Materials Science in Semiconductor Processing}\ }\textbf {\bibinfo
  {volume} {17}},\ \bibinfo {pages} {134 } (\bibinfo {year}
  {2014})}\BibitemShut {NoStop}%
\bibitem [{\citenamefont {Dong}\ \emph {et~al.}(2015)\citenamefont {Dong},
  \citenamefont {Song}, \citenamefont {Wang},\ and\ \citenamefont
  {Wang}}]{Dong_2015}%
  \BibitemOpen
  \bibfield  {author} {\bibinfo {author} {\bibfnamefont {X.}~\bibnamefont
  {Dong}}, \bibinfo {author} {\bibfnamefont {X.}~\bibnamefont {Song}}, \bibinfo
  {author} {\bibfnamefont {Y.}~\bibnamefont {Wang}}, \ and\ \bibinfo {author}
  {\bibfnamefont {J.}~\bibnamefont {Wang}},\ }\href
  {http://stacks.iop.org/1882-0786/8/i=8/a=081302} {\bibfield  {journal}
  {\bibinfo  {journal} {Applied Physics Express}\ }\textbf {\bibinfo {volume}
  {8}},\ \bibinfo {pages} {081302} (\bibinfo {year} {2015})}\BibitemShut
  {NoStop}%
\bibitem [{\citenamefont {Flores}\ and\ \citenamefont
  {Men{\'e}ndez-Proupin}(2016)}]{Flores_2016}%
  \BibitemOpen
  \bibfield  {author} {\bibinfo {author} {\bibfnamefont {M.~A.}\ \bibnamefont
  {Flores}}\ and\ \bibinfo {author} {\bibfnamefont {E.}~\bibnamefont
  {Men{\'e}ndez-Proupin}},\ }\href
  {http://stacks.iop.org/1742-6596/720/i=1/a=012033} {\bibfield  {journal}
  {\bibinfo  {journal} {Journal of Physics: Conference Series}\ }\textbf
  {\bibinfo {volume} {720}},\ \bibinfo {pages} {012033} (\bibinfo {year}
  {2016})}\BibitemShut {NoStop}%
\bibitem [{\citenamefont {Liu}\ \emph {et~al.}(2016)\citenamefont {Liu},
  \citenamefont {Prucnal}, \citenamefont {H{\"u}bner}, \citenamefont {Yuan},
  \citenamefont {Skorupa}, \citenamefont {Helm},\ and\ \citenamefont
  {Zhou}}]{Liu_2016}%
  \BibitemOpen
  \bibfield  {author} {\bibinfo {author} {\bibfnamefont {F.}~\bibnamefont
  {Liu}}, \bibinfo {author} {\bibfnamefont {S.}~\bibnamefont {Prucnal}},
  \bibinfo {author} {\bibfnamefont {R.}~\bibnamefont {H{\"u}bner}}, \bibinfo
  {author} {\bibfnamefont {Y.}~\bibnamefont {Yuan}}, \bibinfo {author}
  {\bibfnamefont {W.}~\bibnamefont {Skorupa}}, \bibinfo {author} {\bibfnamefont
  {M.}~\bibnamefont {Helm}}, \ and\ \bibinfo {author} {\bibfnamefont
  {S.}~\bibnamefont {Zhou}},\ }\href
  {http://stacks.iop.org/0022-3727/49/i=24/a=245104} {\bibfield  {journal}
  {\bibinfo  {journal} {Journal of Physics D: Applied Physics}\ }\textbf
  {\bibinfo {volume} {49}},\ \bibinfo {pages} {245104} (\bibinfo {year}
  {2016})}\BibitemShut {NoStop}%
\bibitem [{\citenamefont {{Carnio}}\ \emph {et~al.}(2017)\citenamefont
  {{Carnio}}, \citenamefont {{Hine}},\ and\ \citenamefont
  {{R{\"o}mer}}}]{Carnio_2017}%
  \BibitemOpen
  \bibfield  {author} {\bibinfo {author} {\bibfnamefont {E.~G.}\ \bibnamefont
  {{Carnio}}}, \bibinfo {author} {\bibfnamefont {N.~D.~M.}\ \bibnamefont
  {{Hine}}}, \ and\ \bibinfo {author} {\bibfnamefont {R.~A.}\ \bibnamefont
  {{R{\"o}mer}}},\ }\href@noop {} {\bibfield  {journal} {\bibinfo  {journal}
  {ArXiv e-prints}\ } (\bibinfo {year} {2017})},\ \Eprint
  {http://arxiv.org/abs/1710.01742} {arXiv:1710.01742} \BibitemShut {NoStop}%
\bibitem [{\citenamefont {Antol{\'i}n}\ \emph {et~al.}(2009)\citenamefont
  {Antol{\'i}n}, \citenamefont {Mart{\'i}}, \citenamefont {Olea}, \citenamefont
  {Pastor}, \citenamefont {Gonz{\'a}lez-D{\'i}az}, \citenamefont {M{\'a}rtil},\
  and\ \citenamefont {Luque}}]{E_Antolin_2009}%
  \BibitemOpen
  \bibfield  {author} {\bibinfo {author} {\bibfnamefont {E.}~\bibnamefont
  {Antol{\'i}n}}, \bibinfo {author} {\bibfnamefont {A.}~\bibnamefont
  {Mart{\'i}}}, \bibinfo {author} {\bibfnamefont {J.}~\bibnamefont {Olea}},
  \bibinfo {author} {\bibfnamefont {D.}~\bibnamefont {Pastor}}, \bibinfo
  {author} {\bibfnamefont {G.}~\bibnamefont {Gonz{\'a}lez-D{\'i}az}}, \bibinfo
  {author} {\bibfnamefont {I.}~\bibnamefont {M{\'a}rtil}}, \ and\ \bibinfo
  {author} {\bibfnamefont {A.}~\bibnamefont {Luque}},\ }\href {\doibase
  10.1063/1.3077202} {\bibfield  {journal} {\bibinfo  {journal} {Appl. Phys.
  Lett.}\ }\textbf {\bibinfo {volume} {94}},\ \bibinfo {pages} {042115}
  (\bibinfo {year} {2009})}\BibitemShut {NoStop}%
\bibitem [{\citenamefont {Berlijn}\ \emph {et~al.}(2011)\citenamefont
  {Berlijn}, \citenamefont {Volja},\ and\ \citenamefont {Ku}}]{Berlijn_2011}%
  \BibitemOpen
  \bibfield  {author} {\bibinfo {author} {\bibfnamefont {T.}~\bibnamefont
  {Berlijn}}, \bibinfo {author} {\bibfnamefont {D.}~\bibnamefont {Volja}}, \
  and\ \bibinfo {author} {\bibfnamefont {W.}~\bibnamefont {Ku}},\ }\href
  {\doibase 10.1103/PhysRevLett.106.077005} {\bibfield  {journal} {\bibinfo
  {journal} {Phys. Rev. Lett.}\ }\textbf {\bibinfo {volume} {106}},\ \bibinfo
  {pages} {077005} (\bibinfo {year} {2011})}\BibitemShut {NoStop}%
\bibitem [{\citenamefont {Ekuma}\ \emph {et~al.}(2014)\citenamefont {Ekuma},
  \citenamefont {Terletska}, \citenamefont {Tam}, \citenamefont {Meng},
  \citenamefont {Moreno},\ and\ \citenamefont {Jarrell}}]{c_ekuma_14}%
  \BibitemOpen
  \bibfield  {author} {\bibinfo {author} {\bibfnamefont {C.~E.}\ \bibnamefont
  {Ekuma}}, \bibinfo {author} {\bibfnamefont {H.}~\bibnamefont {Terletska}},
  \bibinfo {author} {\bibfnamefont {K.-M.}\ \bibnamefont {Tam}}, \bibinfo
  {author} {\bibfnamefont {Z.-Y.}\ \bibnamefont {Meng}}, \bibinfo {author}
  {\bibfnamefont {J.}~\bibnamefont {Moreno}}, \ and\ \bibinfo {author}
  {\bibfnamefont {M.}~\bibnamefont {Jarrell}},\ }\href@noop {} {\bibfield
  {journal} {\bibinfo  {journal} {Phys. Rev. B}\ }\textbf {\bibinfo {volume}
  {89}},\ \bibinfo {pages} {081107} (\bibinfo {year} {2014})}\BibitemShut
  {NoStop}%
\bibitem [{\citenamefont {Vasquez}\ \emph {et~al.}(2008)\citenamefont
  {Vasquez}, \citenamefont {Rodriguez},\ and\ \citenamefont
  {R\"omer}}]{vas_2008}%
  \BibitemOpen
  \bibfield  {author} {\bibinfo {author} {\bibfnamefont {L.~J.}\ \bibnamefont
  {Vasquez}}, \bibinfo {author} {\bibfnamefont {A.}~\bibnamefont {Rodriguez}},
  \ and\ \bibinfo {author} {\bibfnamefont {R.~A.}\ \bibnamefont {R\"omer}},\
  }\href {\doibase 10.1103/PhysRevB.78.195106} {\bibfield  {journal} {\bibinfo
  {journal} {Phys. Rev. B}\ }\textbf {\bibinfo {volume} {78}},\ \bibinfo
  {pages} {195106} (\bibinfo {year} {2008})}\BibitemShut {NoStop}%
\bibitem [{\citenamefont {Zhang}\ \emph {et~al.}(2015)\citenamefont {Zhang},
  \citenamefont {Terletska}, \citenamefont {Moore}, \citenamefont {Ekuma},
  \citenamefont {Tam}, \citenamefont {Berlijn}, \citenamefont {Ku},
  \citenamefont {Moreno},\ and\ \citenamefont {Jarrell}}]{y_zhang_15}%
  \BibitemOpen
  \bibfield  {author} {\bibinfo {author} {\bibfnamefont {Y.}~\bibnamefont
  {Zhang}}, \bibinfo {author} {\bibfnamefont {H.}~\bibnamefont {Terletska}},
  \bibinfo {author} {\bibfnamefont {C.}~\bibnamefont {Moore}}, \bibinfo
  {author} {\bibfnamefont {C.}~\bibnamefont {Ekuma}}, \bibinfo {author}
  {\bibfnamefont {K.-M.}\ \bibnamefont {Tam}}, \bibinfo {author} {\bibfnamefont
  {T.}~\bibnamefont {Berlijn}}, \bibinfo {author} {\bibfnamefont
  {W.}~\bibnamefont {Ku}}, \bibinfo {author} {\bibfnamefont {J.}~\bibnamefont
  {Moreno}}, \ and\ \bibinfo {author} {\bibfnamefont {M.}~\bibnamefont
  {Jarrell}},\ }\href {\doibase 10.1103/PhysRevB.92.205111} {\bibfield
  {journal} {\bibinfo  {journal} {Phys. Rev. B}\ }\textbf {\bibinfo {volume}
  {92}},\ \bibinfo {pages} {205111} (\bibinfo {year} {2015})}\BibitemShut
  {NoStop}%
\bibitem [{\citenamefont {Zhang}\ \emph {et~al.}(2016)\citenamefont {Zhang},
  \citenamefont {Nelson}, \citenamefont {Siddiqui}, \citenamefont {Tam},
  \citenamefont {Yu}, \citenamefont {Berlijn}, \citenamefont {Ku},
  \citenamefont {Vidhyadhiraja}, \citenamefont {Moreno},\ and\ \citenamefont
  {Jarrell}}]{y_zhang_17}%
  \BibitemOpen
  \bibfield  {author} {\bibinfo {author} {\bibfnamefont {Y.}~\bibnamefont
  {Zhang}}, \bibinfo {author} {\bibfnamefont {R.}~\bibnamefont {Nelson}},
  \bibinfo {author} {\bibfnamefont {E.}~\bibnamefont {Siddiqui}}, \bibinfo
  {author} {\bibfnamefont {K.-M.}\ \bibnamefont {Tam}}, \bibinfo {author}
  {\bibfnamefont {U.}~\bibnamefont {Yu}}, \bibinfo {author} {\bibfnamefont
  {T.}~\bibnamefont {Berlijn}}, \bibinfo {author} {\bibfnamefont
  {W.}~\bibnamefont {Ku}}, \bibinfo {author} {\bibfnamefont {N.~S.}\
  \bibnamefont {Vidhyadhiraja}}, \bibinfo {author} {\bibfnamefont
  {J.}~\bibnamefont {Moreno}}, \ and\ \bibinfo {author} {\bibfnamefont
  {M.}~\bibnamefont {Jarrell}},\ }\href {\doibase 10.1103/PhysRevB.94.224208}
  {\bibfield  {journal} {\bibinfo  {journal} {Phys. Rev. B}\ }\textbf {\bibinfo
  {volume} {94}},\ \bibinfo {pages} {224208} (\bibinfo {year}
  {2016})}\BibitemShut {NoStop}%
\bibitem [{\citenamefont {Marzari}\ and\ \citenamefont
  {Vanderbilt}(1997)}]{marzari97_prb56_12847}%
  \BibitemOpen
  \bibfield  {author} {\bibinfo {author} {\bibfnamefont {N.}~\bibnamefont
  {Marzari}}\ and\ \bibinfo {author} {\bibfnamefont {D.}~\bibnamefont
  {Vanderbilt}},\ }\href@noop {} {\bibfield  {journal} {\bibinfo  {journal}
  {Phys. Rev. B}\ }\textbf {\bibinfo {volume} {56}},\ \bibinfo {pages} {12847}
  (\bibinfo {year} {1997})}\BibitemShut {NoStop}%
\bibitem [{\citenamefont {Ku}\ \emph {et~al.}(2002)\citenamefont {Ku},
  \citenamefont {Rosner}, \citenamefont {Pickett},\ and\ \citenamefont
  {Scalettar}}]{w_ku_02}%
  \BibitemOpen
  \bibfield  {author} {\bibinfo {author} {\bibfnamefont {W.}~\bibnamefont
  {Ku}}, \bibinfo {author} {\bibfnamefont {H.}~\bibnamefont {Rosner}}, \bibinfo
  {author} {\bibfnamefont {W.~E.}\ \bibnamefont {Pickett}}, \ and\ \bibinfo
  {author} {\bibfnamefont {R.~T.}\ \bibnamefont {Scalettar}},\ }\href@noop {}
  {\bibfield  {journal} {\bibinfo  {journal} {Phys. Rev. Lett.}\ }\textbf
  {\bibinfo {volume} {89}},\ \bibinfo {pages} {167204} (\bibinfo {year}
  {2002})}\BibitemShut {NoStop}%
\bibitem [{sup()}]{supp}%
  \BibitemOpen
  \href@noop {} {}\bibinfo {note} {Technical details are provided in the
  Supplementary Information.}\BibitemShut {Stop}%
\bibitem [{\citenamefont {Tran}\ and\ \citenamefont {Blaha}(2009)}]{tran_09}%
  \BibitemOpen
  \bibfield  {author} {\bibinfo {author} {\bibfnamefont {F.}~\bibnamefont
  {Tran}}\ and\ \bibinfo {author} {\bibfnamefont {P.}~\bibnamefont {Blaha}},\
  }\href {\doibase 10.1103/PhysRevLett.102.226401} {\bibfield  {journal}
  {\bibinfo  {journal} {Phys. Rev. Lett.}\ }\textbf {\bibinfo {volume} {102}},\
  \bibinfo {pages} {226401} (\bibinfo {year} {2009})}\BibitemShut {NoStop}%
\bibitem [{\citenamefont {Dobrosavljevi\'{c}}\ \emph
  {et~al.}(2003)\citenamefont {Dobrosavljevi\'{c}}, \citenamefont {Pastor},\
  and\ \citenamefont {Nikoli\'{c}}}]{v_dobrosavljevic_03}%
  \BibitemOpen
  \bibfield  {author} {\bibinfo {author} {\bibfnamefont {V.}~\bibnamefont
  {Dobrosavljevi\'{c}}}, \bibinfo {author} {\bibfnamefont {A.~A.}\ \bibnamefont
  {Pastor}}, \ and\ \bibinfo {author} {\bibfnamefont {B.~K.}\ \bibnamefont
  {Nikoli\'{c}}},\ }\href@noop {} {\bibfield  {journal} {\bibinfo  {journal}
  {Europhys. Lett.}\ }\textbf {\bibinfo {volume} {62}},\ \bibinfo {pages} {76}
  (\bibinfo {year} {2003})}\BibitemShut {NoStop}%
\bibitem [{\citenamefont {Soven}(1967)}]{p_soven_67}%
  \BibitemOpen
  \bibfield  {author} {\bibinfo {author} {\bibfnamefont {P.}~\bibnamefont
  {Soven}},\ }\href {\doibase 10.1103/PhysRev.156.809} {\bibfield  {journal}
  {\bibinfo  {journal} {Phys. Rev.}\ }\textbf {\bibinfo {volume} {156}},\
  \bibinfo {pages} {809} (\bibinfo {year} {1967})}\BibitemShut {NoStop}%
\bibitem [{\citenamefont {Wei\ss{}e}\ \emph {et~al.}(2006)\citenamefont
  {Wei\ss{}e}, \citenamefont {Wellein}, \citenamefont {Alvermann},\ and\
  \citenamefont {Fehske}}]{kpm}%
  \BibitemOpen
  \bibfield  {author} {\bibinfo {author} {\bibfnamefont {A.}~\bibnamefont
  {Wei\ss{}e}}, \bibinfo {author} {\bibfnamefont {G.}~\bibnamefont {Wellein}},
  \bibinfo {author} {\bibfnamefont {A.}~\bibnamefont {Alvermann}}, \ and\
  \bibinfo {author} {\bibfnamefont {H.}~\bibnamefont {Fehske}},\ }\href
  {\doibase 10.1103/RevModPhys.78.275} {\bibfield  {journal} {\bibinfo
  {journal} {Rev. Mod. Phys.}\ }\textbf {\bibinfo {volume} {78}},\ \bibinfo
  {pages} {275} (\bibinfo {year} {2006})}\BibitemShut {NoStop}%
\bibitem [{\citenamefont {Terletska}\ \emph {et~al.}(2014)\citenamefont
  {Terletska}, \citenamefont {Ekuma}, \citenamefont {Moore}, \citenamefont
  {Tam}, \citenamefont {Moreno},\ and\ \citenamefont {Jarrell}}]{h_ter_14}%
  \BibitemOpen
  \bibfield  {author} {\bibinfo {author} {\bibfnamefont {H.}~\bibnamefont
  {Terletska}}, \bibinfo {author} {\bibfnamefont {C.~E.}\ \bibnamefont
  {Ekuma}}, \bibinfo {author} {\bibfnamefont {C.}~\bibnamefont {Moore}},
  \bibinfo {author} {\bibfnamefont {K.-M.}\ \bibnamefont {Tam}}, \bibinfo
  {author} {\bibfnamefont {J.}~\bibnamefont {Moreno}}, \ and\ \bibinfo {author}
  {\bibfnamefont {M.}~\bibnamefont {Jarrell}},\ }\href {\doibase
  10.1103/PhysRevB.90.094208} {\bibfield  {journal} {\bibinfo  {journal} {Phys.
  Rev. B}\ }\textbf {\bibinfo {volume} {90}},\ \bibinfo {pages} {094208}
  (\bibinfo {year} {2014})}\BibitemShut {NoStop}%
\bibitem [{\citenamefont {Rosenbaum}\ \emph {et~al.}(1983)\citenamefont
  {Rosenbaum}, \citenamefont {Milligan}, \citenamefont {Paalanen},
  \citenamefont {Thomas}, \citenamefont {Bhatt},\ and\ \citenamefont
  {Lin}}]{Rosenbaum_1983}%
  \BibitemOpen
  \bibfield  {author} {\bibinfo {author} {\bibfnamefont {T.~F.}\ \bibnamefont
  {Rosenbaum}}, \bibinfo {author} {\bibfnamefont {R.~F.}\ \bibnamefont
  {Milligan}}, \bibinfo {author} {\bibfnamefont {M.~A.}\ \bibnamefont
  {Paalanen}}, \bibinfo {author} {\bibfnamefont {G.~A.}\ \bibnamefont
  {Thomas}}, \bibinfo {author} {\bibfnamefont {R.~N.}\ \bibnamefont {Bhatt}}, \
  and\ \bibinfo {author} {\bibfnamefont {W.}~\bibnamefont {Lin}},\ }\href
  {\doibase 10.1103/PhysRevB.27.7509} {\bibfield  {journal} {\bibinfo
  {journal} {Phys. Rev. B}\ }\textbf {\bibinfo {volume} {27}},\ \bibinfo
  {pages} {7509} (\bibinfo {year} {1983})}\BibitemShut {NoStop}%
\bibitem [{\citenamefont {Lucena}\ \emph {et~al.}(2008)\citenamefont {Lucena},
  \citenamefont {Aguilera}, \citenamefont {Palacios}, \citenamefont
  {Wahn{\'o}n},\ and\ \citenamefont {Conesa}}]{Raquel_2008}%
  \BibitemOpen
  \bibfield  {author} {\bibinfo {author} {\bibfnamefont {R.}~\bibnamefont
  {Lucena}}, \bibinfo {author} {\bibfnamefont {I.}~\bibnamefont {Aguilera}},
  \bibinfo {author} {\bibfnamefont {P.}~\bibnamefont {Palacios}}, \bibinfo
  {author} {\bibfnamefont {P.}~\bibnamefont {Wahn{\'o}n}}, \ and\ \bibinfo
  {author} {\bibfnamefont {J.~C.}\ \bibnamefont {Conesa}},\ }\href {\doibase
  10.1021/cm801128b} {\bibfield  {journal} {\bibinfo  {journal} {Chemistry of
  Materials}\ }\textbf {\bibinfo {volume} {20}},\ \bibinfo {pages} {5125}
  (\bibinfo {year} {2008})}\BibitemShut {NoStop}%
\bibitem [{\citenamefont {Brandt}\ \emph {et~al.}(1989)\citenamefont {Brandt},
  \citenamefont {Hennel}, \citenamefont {Bryskiewicz}, \citenamefont {Ko},
  \citenamefont {Pawlosicz},\ and\ \citenamefont {Gatos}}]{Brandt_1989}%
  \BibitemOpen
  \bibfield  {author} {\bibinfo {author} {\bibfnamefont {C.~D.}\ \bibnamefont
  {Brandt}}, \bibinfo {author} {\bibfnamefont {A.~M.}\ \bibnamefont {Hennel}},
  \bibinfo {author} {\bibfnamefont {T.}~\bibnamefont {Bryskiewicz}}, \bibinfo
  {author} {\bibfnamefont {K.~Y.}\ \bibnamefont {Ko}}, \bibinfo {author}
  {\bibfnamefont {L.~M.}\ \bibnamefont {Pawlosicz}}, \ and\ \bibinfo {author}
  {\bibfnamefont {H.~C.}\ \bibnamefont {Gatos}},\ }\href {\doibase
  10.1063/1.342614} {\bibfield  {journal} {\bibinfo  {journal} {Journal of
  Applied Physics}\ }\textbf {\bibinfo {volume} {65}},\ \bibinfo {pages} {3459}
  (\bibinfo {year} {1989})}\BibitemShut {NoStop}%
\bibitem [{\citenamefont {Olsson}\ \emph {et~al.}(2009)\citenamefont {Olsson},
  \citenamefont {Domain},\ and\ \citenamefont {Guillemoles}}]{Olsson_2009}%
  \BibitemOpen
  \bibfield  {author} {\bibinfo {author} {\bibfnamefont {P.}~\bibnamefont
  {Olsson}}, \bibinfo {author} {\bibfnamefont {C.}~\bibnamefont {Domain}}, \
  and\ \bibinfo {author} {\bibfnamefont {J.-F.}\ \bibnamefont {Guillemoles}},\
  }\href {\doibase 10.1103/PhysRevLett.102.227204} {\bibfield  {journal}
  {\bibinfo  {journal} {Phys. Rev. Lett.}\ }\textbf {\bibinfo {volume} {102}},\
  \bibinfo {pages} {227204} (\bibinfo {year} {2009})}\BibitemShut {NoStop}%
\bibitem [{\citenamefont {Abrahams}(2010)}]{abrahams_2014}%
  \BibitemOpen
  \bibfield  {author} {\bibinfo {author} {\bibfnamefont {E.}~\bibnamefont
  {Abrahams}},\ }\href@noop {} {\emph {\bibinfo {title} {50 Years of Anderson
  Localization}}}\ (\bibinfo  {publisher} {World Scientific},\ \bibinfo {year}
  {2010})\BibitemShut {NoStop}%
\bibitem [{\citenamefont {Chopra}\ and\ \citenamefont
  {Bahl}(1970)}]{chopra_1970}%
  \BibitemOpen
  \bibfield  {author} {\bibinfo {author} {\bibfnamefont {K.~L.}\ \bibnamefont
  {Chopra}}\ and\ \bibinfo {author} {\bibfnamefont {S.~K.}\ \bibnamefont
  {Bahl}},\ }\href {\doibase 10.1103/PhysRevB.1.2545} {\bibfield  {journal}
  {\bibinfo  {journal} {Phys. Rev. B}\ }\textbf {\bibinfo {volume} {1}},\
  \bibinfo {pages} {2545} (\bibinfo {year} {1970})}\BibitemShut {NoStop}%
\bibitem [{\citenamefont {Bates}\ and\ \citenamefont
  {Zhang}(2013)}]{Clayton_2013}%
  \BibitemOpen
  \bibfield  {author} {\bibinfo {author} {\bibfnamefont {C.~W.}\ \bibnamefont
  {Bates}}\ and\ \bibinfo {author} {\bibfnamefont {C.}~\bibnamefont {Zhang}},\
  }\href {\doibase 10.1063/1.4824854} {\bibfield  {journal} {\bibinfo
  {journal} {AIP Advances}\ }\textbf {\bibinfo {volume} {3}},\ \bibinfo {pages}
  {102111} (\bibinfo {year} {2013})}\BibitemShut {NoStop}%
\bibitem [{\citenamefont {van Hapert}(1973)}]{hapert_1973}%
  \BibitemOpen
  \bibfield  {author} {\bibinfo {author} {\bibfnamefont {J.~J.}\ \bibnamefont
  {van Hapert}},\ }\emph {\bibinfo {title} {Hopping Conduction and Chemical
  Structure: a study on Silicon Suboxides}},\ \href@noop {} {Ph.D. thesis},\
  \bibinfo  {school} {Utrecht University} (\bibinfo {year} {1973})\BibitemShut
  {NoStop}%
\bibitem [{\citenamefont {Ambegaokar}\ \emph {et~al.}(1971)\citenamefont
  {Ambegaokar}, \citenamefont {Halperin},\ and\ \citenamefont
  {Langer}}]{Ambe_1971}%
  \BibitemOpen
  \bibfield  {author} {\bibinfo {author} {\bibfnamefont {V.}~\bibnamefont
  {Ambegaokar}}, \bibinfo {author} {\bibfnamefont {B.~I.}\ \bibnamefont
  {Halperin}}, \ and\ \bibinfo {author} {\bibfnamefont {J.~S.}\ \bibnamefont
  {Langer}},\ }\href {\doibase 10.1103/PhysRevB.4.2612} {\bibfield  {journal}
  {\bibinfo  {journal} {Phys. Rev. B}\ }\textbf {\bibinfo {volume} {4}},\
  \bibinfo {pages} {2612} (\bibinfo {year} {1971})}\BibitemShut {NoStop}%
\end{thebibliography}

\begin{thebibliography}{8}%
\makeatletter
\providecommand \@ifxundefined [1]{%
 \@ifx{#1\undefined}
}%
\providecommand \@ifnum [1]{%
 \ifnum #1\expandafter \@firstoftwo
 \else \expandafter \@secondoftwo
 \fi
}%
\providecommand \@ifx [1]{%
 \ifx #1\expandafter \@firstoftwo
 \else \expandafter \@secondoftwo
 \fi
}%
\providecommand \natexlab [1]{#1}%
\providecommand \enquote  [1]{``#1''}%
\providecommand \bibnamefont  [1]{#1}%
\providecommand \bibfnamefont [1]{#1}%
\providecommand \citenamefont [1]{#1}%
\providecommand \href@noop [0]{\@secondoftwo}%
\providecommand \href [0]{\begingroup \@sanitize@url \@href}%
\providecommand \@href[1]{\@@startlink{#1}\@@href}%
\providecommand \@@href[1]{\endgroup#1\@@endlink}%
\providecommand \@sanitize@url [0]{\catcode `\\12\catcode `\$12\catcode
  `\&12\catcode `\#12\catcode `\^12\catcode `\_12\catcode `\%12\relax}%
\providecommand \@@startlink[1]{}%
\providecommand \@@endlink[0]{}%
\providecommand \url  [0]{\begingroup\@sanitize@url \@url }%
\providecommand \@url [1]{\endgroup\@href {#1}{\urlprefix }}%
\providecommand \urlprefix  [0]{URL }%
\providecommand \Eprint [0]{\href }%
\providecommand \doibase [0]{http://dx.doi.org/}%
\providecommand \selectlanguage [0]{\@gobble}%
\providecommand \bibinfo  [0]{\@secondoftwo}%
\providecommand \bibfield  [0]{\@secondoftwo}%
\providecommand \translation [1]{[#1]}%
\providecommand \BibitemOpen [0]{}%
\providecommand \bibitemStop [0]{}%
\providecommand \bibitemNoStop [0]{.\EOS\space}%
\providecommand \EOS [0]{\spacefactor3000\relax}%
\providecommand \BibitemShut  [1]{\csname bibitem#1\endcsname}%
\let\auto@bib@innerbib\@empty
\bibitem [{\citenamefont {Schwarz}\ \emph {et~al.}(2002)\citenamefont
  {Schwarz}, \citenamefont {Blaha},\ and\ \citenamefont {Madsen}}]{wien2k}%
  \BibitemOpen
  \bibfield  {author} {\bibinfo {author} {\bibfnamefont {K.}~\bibnamefont
  {Schwarz}}, \bibinfo {author} {\bibfnamefont {P.}~\bibnamefont {Blaha}}, \
  and\ \bibinfo {author} {\bibfnamefont {G.}~\bibnamefont {Madsen}},\ }\href
  {\doibase https://doi.org/10.1016/S0010-4655(02)00206-0} {\bibfield
  {journal} {\bibinfo  {journal} {Computer Physics Communications}\ }\textbf
  {\bibinfo {volume} {147}},\ \bibinfo {pages} {71 } (\bibinfo {year}
  {2002})},\ \bibinfo {note} {proceedings of the Europhysics Conference on
  Computational Physics Computational Modeling and Simulation of Complex
  Systems}\BibitemShut {NoStop}%
\bibitem [{\citenamefont {Perdew}\ \emph {et~al.}(1996)\citenamefont {Perdew},
  \citenamefont {Burke},\ and\ \citenamefont {Ernzerhof}}]{perdew_1996}%
  \BibitemOpen
  \bibfield  {author} {\bibinfo {author} {\bibfnamefont {J.~P.}\ \bibnamefont
  {Perdew}}, \bibinfo {author} {\bibfnamefont {K.}~\bibnamefont {Burke}}, \
  and\ \bibinfo {author} {\bibfnamefont {M.}~\bibnamefont {Ernzerhof}},\ }\href
  {\doibase 10.1103/PhysRevLett.77.3865} {\bibfield  {journal} {\bibinfo
  {journal} {Phys. Rev. Lett.}\ }\textbf {\bibinfo {volume} {77}},\ \bibinfo
  {pages} {3865} (\bibinfo {year} {1996})}\BibitemShut {NoStop}%
\bibitem [{\citenamefont {T{\"{o}}bbens}\ \emph {et~al.}(2001)\citenamefont
  {T{\"{o}}bbens}, \citenamefont {St{\"{u}}{\ss }er}, \citenamefont {Knorr},
  \citenamefont {Mayer},\ and\ \citenamefont {Lampert}}]{tobbens2001}%
  \BibitemOpen
  \bibfield  {author} {\bibinfo {author} {\bibfnamefont {D.}~\bibnamefont
  {T{\"{o}}bbens}}, \bibinfo {author} {\bibfnamefont {N.}~\bibnamefont
  {St{\"{u}}{\ss }er}}, \bibinfo {author} {\bibfnamefont {K.}~\bibnamefont
  {Knorr}}, \bibinfo {author} {\bibfnamefont {H.}~\bibnamefont {Mayer}}, \ and\
  \bibinfo {author} {\bibfnamefont {G.}~\bibnamefont {Lampert}},\ }in\ \href
  {\doibase 10.4028/www.scientific.net/MSF.378-381.288} {\emph {\bibinfo
  {booktitle} {European Powder Diffraction EPDIC 7}}},\ \bibinfo {series}
  {Materials Science Forum}, Vol.\ \bibinfo {volume} {378}\ (\bibinfo
  {publisher} {Trans Tech Publications},\ \bibinfo {year} {2001})\ pp.\
  \bibinfo {pages} {288--293}\BibitemShut {NoStop}%
\bibitem [{\citenamefont {S\'anchez}\ \emph {et~al.}(2009)\citenamefont
  {S\'anchez}, \citenamefont {Aguilera}, \citenamefont {Palacios},\ and\
  \citenamefont {Wahn\'on}}]{K_Sanchez_2009_supp}%
  \BibitemOpen
  \bibfield  {author} {\bibinfo {author} {\bibfnamefont {K.}~\bibnamefont
  {S\'anchez}}, \bibinfo {author} {\bibfnamefont {I.}~\bibnamefont {Aguilera}},
  \bibinfo {author} {\bibfnamefont {P.}~\bibnamefont {Palacios}}, \ and\
  \bibinfo {author} {\bibfnamefont {P.}~\bibnamefont {Wahn\'on}},\ }\href
  {\doibase 10.1103/PhysRevB.79.165203} {\bibfield  {journal} {\bibinfo
  {journal} {Phys. Rev. B}\ }\textbf {\bibinfo {volume} {79}},\ \bibinfo
  {pages} {165203} (\bibinfo {year} {2009})}\BibitemShut {NoStop}%
\bibitem [{\citenamefont {Anisimov}\ \emph {et~al.}(1993)\citenamefont
  {Anisimov}, \citenamefont {Solovyev}, \citenamefont {Korotin}, \citenamefont
  {Czy\ifmmode~\dot{z}\else \.{z}\fi{}yk},\ and\ \citenamefont
  {Sawatzky}}]{anisimov_1993}%
  \BibitemOpen
  \bibfield  {author} {\bibinfo {author} {\bibfnamefont {V.~I.}\ \bibnamefont
  {Anisimov}}, \bibinfo {author} {\bibfnamefont {I.~V.}\ \bibnamefont
  {Solovyev}}, \bibinfo {author} {\bibfnamefont {M.~A.}\ \bibnamefont
  {Korotin}}, \bibinfo {author} {\bibfnamefont {M.~T.}\ \bibnamefont
  {Czy\ifmmode~\dot{z}\else \.{z}\fi{}yk}}, \ and\ \bibinfo {author}
  {\bibfnamefont {G.~A.}\ \bibnamefont {Sawatzky}},\ }\href {\doibase
  10.1103/PhysRevB.48.16929} {\bibfield  {journal} {\bibinfo  {journal} {Phys.
  Rev. B}\ }\textbf {\bibinfo {volume} {48}},\ \bibinfo {pages} {16929}
  (\bibinfo {year} {1993})}\BibitemShut {NoStop}%
\bibitem [{\citenamefont {Tran}\ and\ \citenamefont
  {Blaha}(2009)}]{tran_09_supp}%
  \BibitemOpen
  \bibfield  {author} {\bibinfo {author} {\bibfnamefont {F.}~\bibnamefont
  {Tran}}\ and\ \bibinfo {author} {\bibfnamefont {P.}~\bibnamefont {Blaha}},\
  }\href {\doibase 10.1103/PhysRevLett.102.226401} {\bibfield  {journal}
  {\bibinfo  {journal} {Phys. Rev. Lett.}\ }\textbf {\bibinfo {volume} {102}},\
  \bibinfo {pages} {226401} (\bibinfo {year} {2009})}\BibitemShut {NoStop}%
\bibitem [{\citenamefont {Ku}\ \emph {et~al.}(2002)\citenamefont {Ku},
  \citenamefont {Rosner}, \citenamefont {Pickett},\ and\ \citenamefont
  {Scalettar}}]{w_ku_02_supp}%
  \BibitemOpen
  \bibfield  {author} {\bibinfo {author} {\bibfnamefont {W.}~\bibnamefont
  {Ku}}, \bibinfo {author} {\bibfnamefont {H.}~\bibnamefont {Rosner}}, \bibinfo
  {author} {\bibfnamefont {W.~E.}\ \bibnamefont {Pickett}}, \ and\ \bibinfo
  {author} {\bibfnamefont {R.~T.}\ \bibnamefont {Scalettar}},\ }\href@noop {}
  {\bibfield  {journal} {\bibinfo  {journal} {Phys. Rev. Lett.}\ }\textbf
  {\bibinfo {volume} {89}},\ \bibinfo {pages} {167204} (\bibinfo {year}
  {2002})}\BibitemShut {NoStop}%
\bibitem [{\citenamefont {Jarrell}\ and\ \citenamefont
  {Krishnamurthy}(2001)}]{jarrell_2001}%
  \BibitemOpen
  \bibfield  {author} {\bibinfo {author} {\bibfnamefont {M.}~\bibnamefont
  {Jarrell}}\ and\ \bibinfo {author} {\bibfnamefont {H.~R.}\ \bibnamefont
  {Krishnamurthy}},\ }\href {\doibase 10.1103/PhysRevB.63.125102} {\bibfield
  {journal} {\bibinfo  {journal} {Phys. Rev. B}\ }\textbf {\bibinfo {volume}
  {63}},\ \bibinfo {pages} {125102} (\bibinfo {year} {2001})}\BibitemShut
  {NoStop}%
\end{thebibliography}
%

\clearpage
\onecolumngrid
\begin{center}
\textbf{\large Supplemental Materials: On the Nature of Localization in Ti doped Si}
\end{center}

\setcounter{equation}{0}
\setcounter{figure}{0}
\setcounter{table}{0}
\setcounter{page}{1}
\makeatletter
\renewcommand{\theequation}{S\arabic{equation}}
\renewcommand{\thefigure}{S\arabic{figure}}
\renewcommand{\bibnumfmt}[1]{[S#1]}
\renewcommand{\citenumfont}[1]{S#1}


\clearpage
\section{1. DETAILS OF THE DENSITY FUNCTIONAL THEORY CALCULATIONS}
\begin{figure}[h!]
 \includegraphics[trim = 0mm 0mm 0mm 0mm,width=0.7\columnwidth,clip=true]{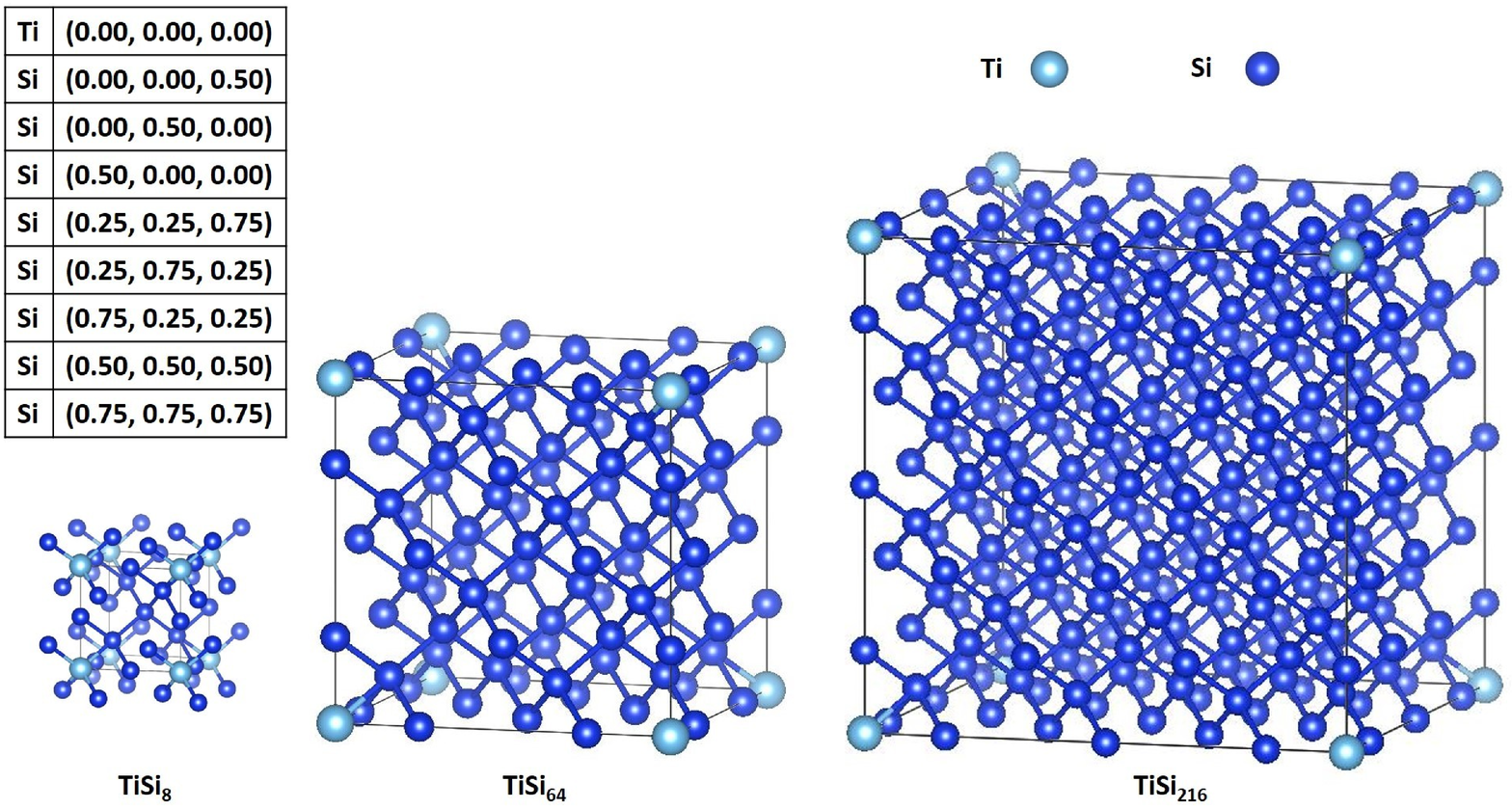}
 \caption{Three supercells used to capture the single Ti impurity influence. For TiSi$_8$ the Ti-Si bonds outside the supercell are shown to illustrate the tetrahedral coordination of the Ti interstitial. The table shows the coordinates of the non-relaxed Ti and Si positions expressed in the supercell lattice vectors. }
 \label{fig:X1}
\end{figure}

\begin{figure}[h!]
 \includegraphics[trim = 0mm 0mm 0mm 0mm,width=0.7\columnwidth,clip=true]{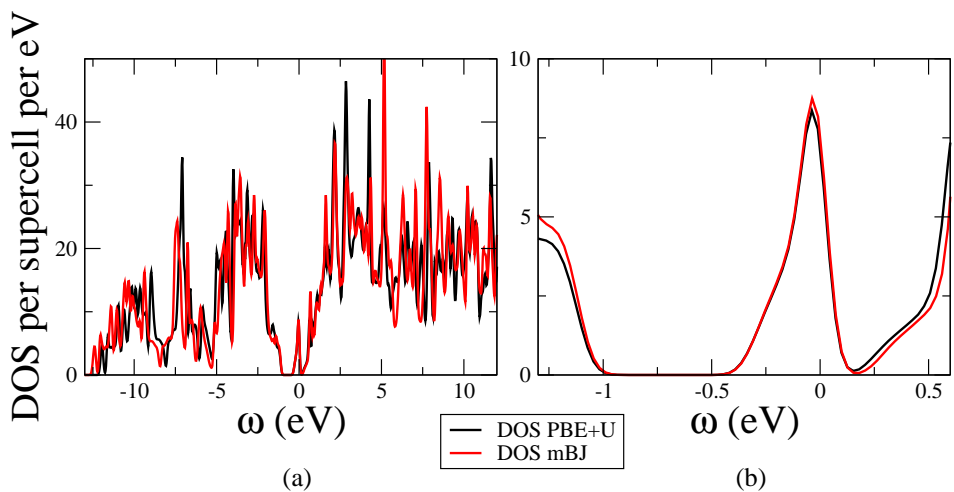}
 \caption{Comparison of Density of States (DOS) of TiSi$_{64}$ obtained from PBE+U (black) and the modified Becke-Johnson potential (mBJ) (red). (a) in a large frequency range, (b) around the impurity band region.}
 \label{fig:X2}
\end{figure}

We used the WIEN2K~\cite{wien2k} implementation of the full potential linearized augmented plane wave (LAPW) method with the Perdew-Burke-Ernzerhof (PBE) exchange-correlation functional~\cite{perdew_1996}. For the simulation we took the space group 227: Fd-3m and the lattice constant a = 10.26 Bohr of Si from Ref.~\cite{tobbens2001}. To capture the single Ti impurity influence, three different supercells were used corresponding to the sizes of the 1 $\times$ 1 $\times$ 1, 2 $\times$ 2 $\times$ 2, 3 $\times$ 3 $\times$ 3 Si$_8$ supercell (see Fig. \ref{fig:X1}). Following the conclusion of Ref.~\cite{K_Sanchez_2009_supp} the Ti is located at the tetrahedral interstitial site (Fig.~\ref{fig:X1}). We use a k-point mesh of 11 $\times$ 11 $\times$ 11 for the undoped normal cell and of 3 $\times$ 3 $\times$ 3 for the  supercells. The basis set sizes were determined by $RK_{max}=6$. To better describe the experimental band-gap of Si we apply the LDA+U approximation~\cite{anisimov_1993}, or to be more precise the PBE+U approximation, with U=-5.1 eV for Si-$p$. As shown in Fig.~\ref{fig:X2} our PBE+U results compare accurately with the results obtained from the modified Becke-Johnson potential~\cite{tran_09_supp}, especially for the Density of States (DOS) close to the gap. For the results shown in Sec. 7
the internal forces were relaxed to less than 2mRy/Bohr. For all other results in the supplement and the manuscript we did not relax the atomic positions.

\clearpage
\section{2. DETAILS OF THE WANNIER FUNCTION CALCULATIONS}

\begin{figure}[h!]
 \includegraphics[trim = 0mm 0mm 0mm 0mm,width=0.9\columnwidth,clip=true]{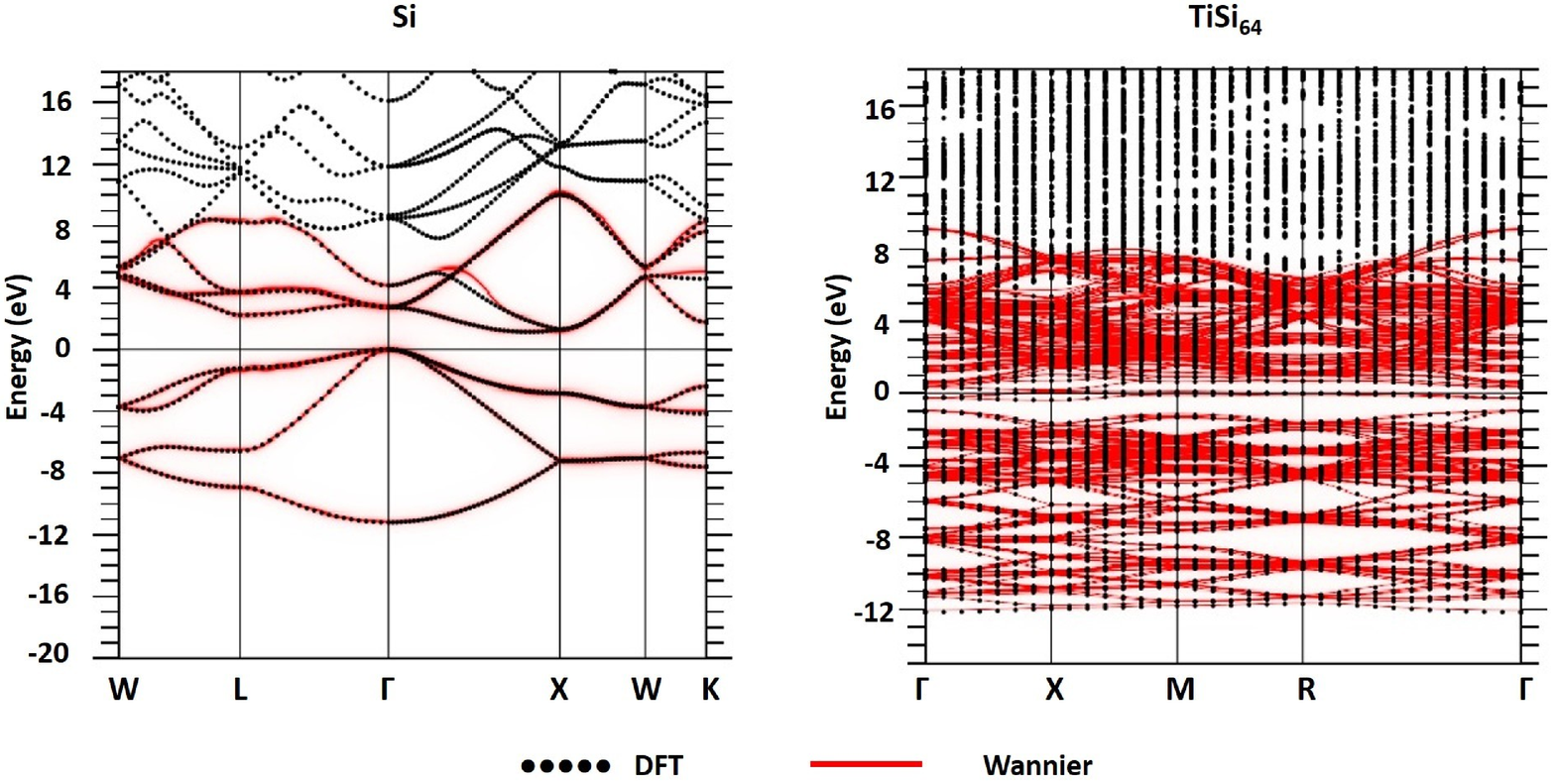}
 \caption{Comparison of Wannier and DFT bandstructure for Si (left) and TiSi$_{64}$ (right).}
 \label{fig:X3}
\end{figure}

To derive the Wannier function based tight binding Hamiltonians from the DFT calculations we perform a projected Wannier function transformation~\cite{w_ku_02_supp}. Specifically we project the Ti-$d$, Si-$s$ and Si-$p$ orbitals onto the bands within [-12.5,11] eV. Fig.~\ref{fig:X3} shows the comparison of the Wannier and DFT bandstructures for pure Si and the TiSi$_{64}$ supercell.  

\section{3. Full DOS spectrum}
\begin{figure}[h!]
 \includegraphics[trim = 0mm 0mm 0mm 0mm,width=0.5\columnwidth,clip=true]{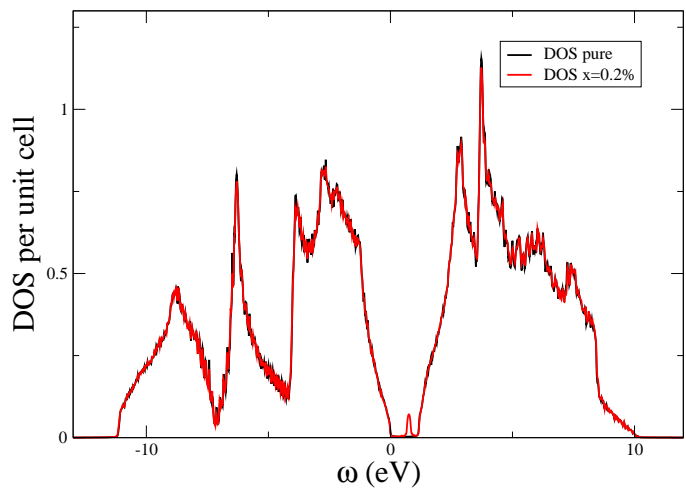}
 \caption{DOS of Si(Ti) with Ti concentration of $x=0.2\%$  (calculated with DCA) compared with the DOS of pure Si.}
 \label{fig:dos_comp}
\end{figure}

The DOS spectrum obtained from DCA~\cite{jarrell_2001} for the undoped case and the Ti concentration $x=0.2\%$ are shown in Fig.~\ref{fig:dos_comp}. We note that the impurity band is located inside the band gap for the doped case.

\clearpage
\section{4. Convergence of DOS and TDOS with cluster size}
\begin{figure}[h!]
 \includegraphics[trim = 0mm 0mm 0mm 0mm,width=0.5\columnwidth,clip=true]{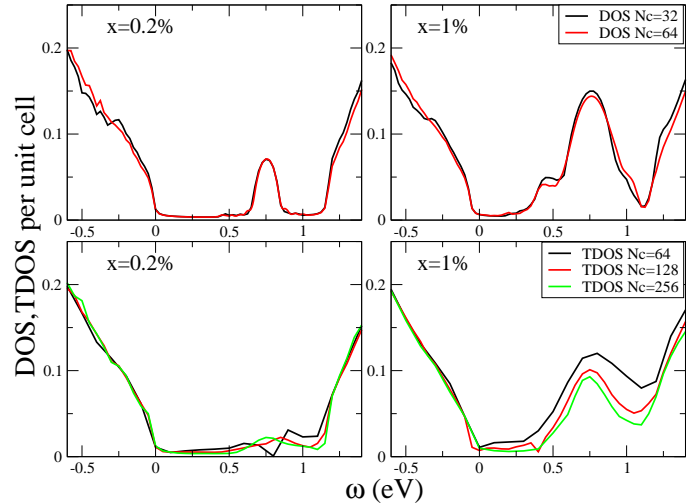}
 \caption{The DOS (upper panel) and TDOS (lower panel) of Si(Ti) with $x=0.2\%$ and $x=1\%$ for various cluster sizes. }
 \label{fig:dos_tdos_Nc}
\end{figure}

We check the convergence of the DCA and the TMDCA with cluster size N$_c$ and find that the DOS converges up to cluster size N$_c$=64 and the TDOS converges up to cluster size N$_c$=128 as shown in Fig.~\ref{fig:dos_tdos_Nc}. We use these cluster sizes throughout the manuscript and the supplement. 

\section{5. Convergence of DOS and TDOS with $k_{mesh}$ in the Brillouin Zone}
\begin{figure}[h!]
 \includegraphics[trim = 0mm 0mm 0mm 0mm,width=0.5\columnwidth,clip=true]{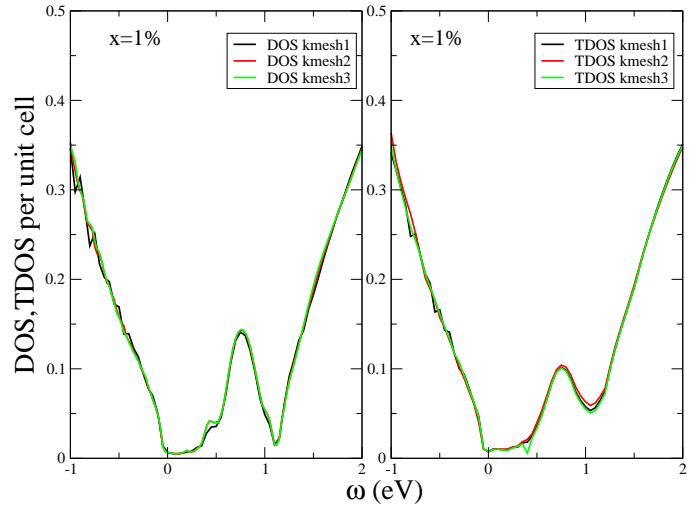}
 \caption{DOS (left) and TDOS (right) for Ti concentration $x=1\%$ and three different values of $k_{mesh}$. kmesh1$=1.08\times 10^5$ k-points, kmesh2$=2.56 \times 10^5$ k-points, kmesh3$=5 \times 10^5$ k-points in BZ. }
 \label{fig:dos_tdos_kmesh}
\end{figure}

We check the convergence of DCA and TMDCA against the 
size of the mesh in momentum space ($k_{mesh}$), and find that both  DOS and TDOS are well converged with $k_{mesh}=5 \times 10^5$ k-points in the Brillouin zone as shown in Fig.~\ref{fig:dos_tdos_kmesh}. We use this number of k-points throughout the manuscript and the supplement.

\clearpage
\section{6. Convergence of DOS and TDOS with R$_{mt}$*K$_{max}$}

\begin{figure}[h!]
 \includegraphics[trim = 0mm 0mm 0mm 0mm,width=0.5\columnwidth,clip=true]{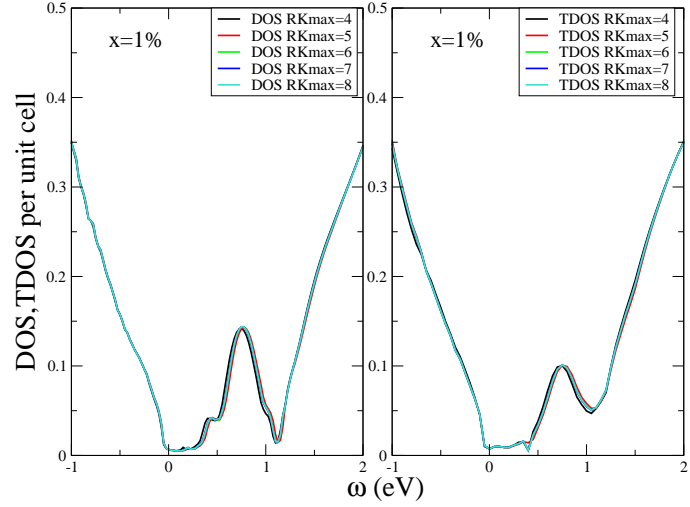}
 \caption{DOS (N$_c$=64) and TDOS (N$_c$=128) with Ti concentration $x=1\%$ for various values of R$_{mt}$*K$_{max}$. }
 \label{fig:dos_rkmax}
\end{figure}

We check the convergence of DOS and TDOS against the number of LAPW basis functions used in the DFT, which is controlled by the parameter R$_{mt}$*K$_{max}$. Specifically we perform DFT calculations of pure Si and TiSi$_{216}$ supercell for various values of R$_{mt}$*K$_{max}$ and derive the impurity potential for each of them. We find that both DOS and TDOS are well converged for R$_{mt}$*K$_{max}$=6 as shown in Fig.~\ref{fig:dos_rkmax}.
We use this value throughout the manuscript and the supplement.

\clearpage

\section{7. Effect of lattice relaxation on DOS and TDOS}\label{sec:relax}
\begin{figure}[h!]
 \includegraphics[trim = 0mm 0mm 0mm 0mm,width=0.5\columnwidth,clip=true]{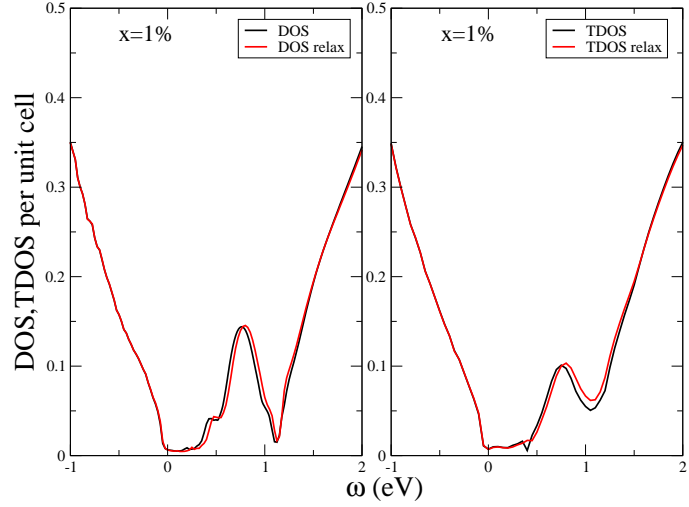}
 \caption{DOS and TDOS with Ti concentration $x=1\%$ based on the impurity potential derived with and without lattice relaxation.  }
 \label{fig:dos_relax}
\end{figure}

\begin{figure}[h!]
 \includegraphics[trim = 0mm 0mm 0mm 0mm,width=0.5\columnwidth,clip=true]{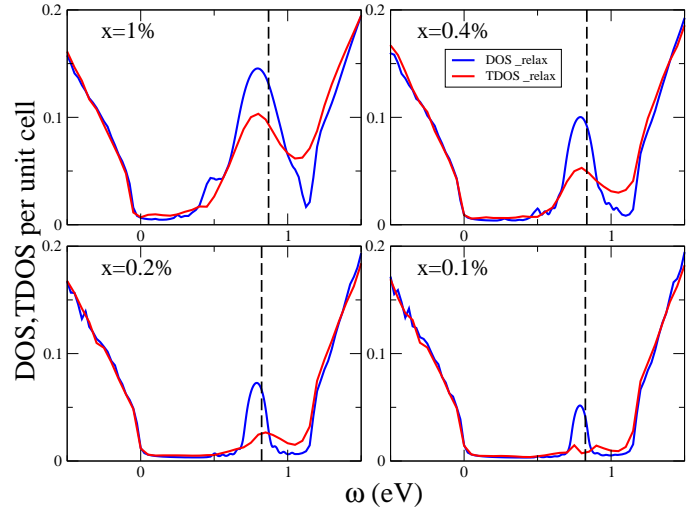}
 \caption{DOS and TDOS of Ti doped Si based on the impurity potential derived using lattice relaxation for 
various Ti concentrations: $x=$1\%, 0.4\%, 0.2\%, 0.1\%. The chemical potential is represented by the dashed line.}
 \label{fig:dos_tdos_<_x_np_relax}
\end{figure}

We derive the impurity potential from the TiSi$_{216}$ supercell using lattice relaxed atomic positions and use it for the DCA and TMDCA calculation. We compare our results with the case without lattice relaxation. The change in DOS and TDOS is negligible as shown in Fig.~\ref{fig:dos_relax}. The critical Ti concentration determined using this impurity potential still lies between 0.1\% and 0.2\% as shown in Fig.~\ref{fig:dos_tdos_<_x_np_relax}.

\clearpage

\pagebreak
\section{8. Effect of spin-polarization on DOS and TDOS}
\begin{figure}[h!]
 \includegraphics[trim = 0mm 0mm 0mm 0mm,width=0.5\columnwidth,clip=true]{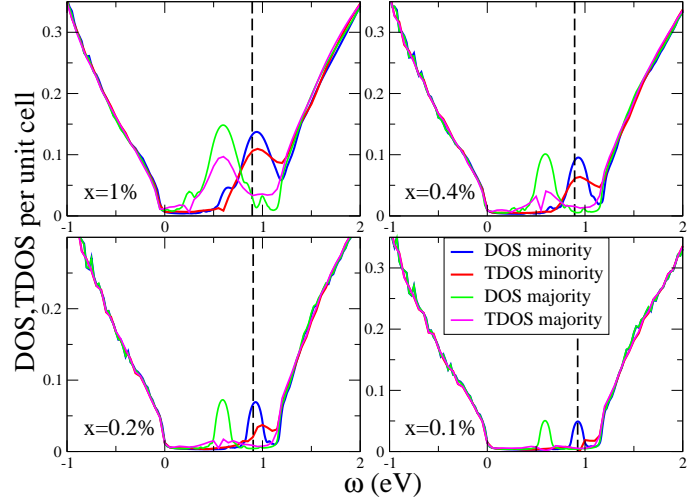}
 \caption{DOS and TDOS of spin-polarized Ti doped Si for 
various Ti concentrations: $x=$1\%, 0.4\%, 0.2\%, 0.1\%. Results for both spin species are plotted. The chemical potential is represented by the dashed line. }
 \label{fig:dos_tdos_x_sp}
\end{figure}

We derive the impurity potential from the TiSi$_{216}$ supercell using spin-polarized DFT and use it for the DCA and TMDCA calculation.  As shown in Fig.~\ref{fig:dos_tdos_x_sp} for Ti concentrations $x<0.4\%$, the spin majority part of the impurity band is fully filled and energetically separated from the partially filled spin minority impurity band. Therefore electrons from the valence band can only be promoted to the spin-minority impurity band and we need focus our study of localization on those spin minority states around the chemical potential. Fig.~\ref{fig:dos_tdos_x_sp} shows that the DOS and TDOS of the spin-minority impurity band follows the same trend as those of the non spin-polarized impurity band presented in Fig. 2 of the manuscript. In particular the critical Ti concentration still lies between 0.1\% and 0.2\%.

\clearpage
\section{9. Comparing DOS and TDOS for TiSi$_{64}$ and TiSi$_{216}$ derived impurity potentials}

\begin{figure}[h!]
 \includegraphics[trim = 0mm 0mm 0mm 0mm,width=0.5\columnwidth,clip=true]{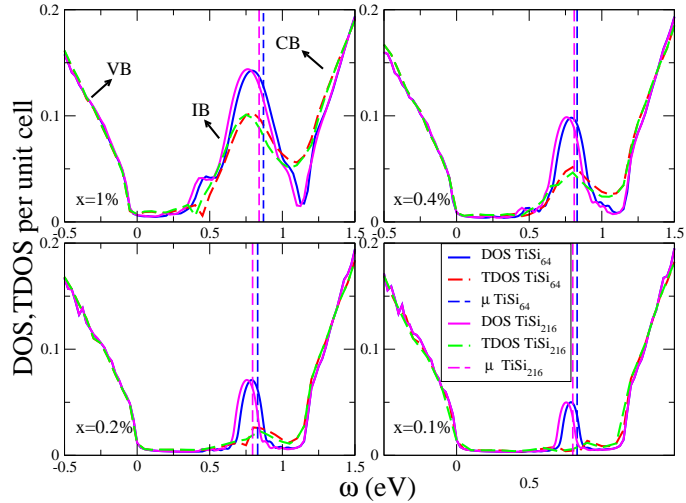}
 \caption{DOS and TDOS of Ti doped Si based on the impurity potential derived from the supercells TiSi$_{64}$ and TiSi$_{216}$, and various Ti concentrations: $x=$1\%, 0.4\%, 0.2\%, 0.1\%. VB, CB, and IB correspond to the valence, conduction  and intermediate band, respectively. The chemical potential is represented by the dashed line. }
 \label{fig:dos_tdos_64_216}
\end{figure}

Fig.~\ref{fig:dos_tdos_64_216} shows that DOS and TDOS change little when the impurity potential is derived from the TiSi$_{64}$ supercell instead  of the TiSi$_{216}$.

\end{document}